\documentclass[hidelinks,a4paper,10pt,onecolumn, switch]{article}
\pdfoutput=1

\usepackage{fancyhdr}
\fancypagestyle{plain}{%
  \fancyhf{}
\fancyfoot[R]{\thepage}
\fancyfoot[L]{\small \textit{Preprint}\\ \texttt{arXiv}}
}

\usepackage[affil-it]{authblk}
\makeatletter
\def\@maketitle{%
  \newpage
  \null
  \vskip 2em%
  \begin{center}%
  \let \footnote \thanks
    {\Large\bfseries \@title \par}%
    \vskip 1.5em%
    {\normalsize
      \lineskip .5em%
      \begin{tabular}[t]{c}%
        \@author
      \end{tabular}\par}%
    \vskip 1em%
    {\normalsize \@date}%
  \end{center}%
  \par
  \vskip 1.5em}
\makeatother

\usepackage[a4paper, total={17.5cm,24cm}]{geometry}

\usepackage[font=normal,labelfont=bf]{caption}

\usepackage{sectsty} %small fonts for appendix by means of \sectionfont{\small}

\usepackage{titlesec}
\titlespacing\section{0pt}{12pt plus 3pt minus 3pt}{1pt plus 1pt minus 1pt}
\titlespacing\subsection{0pt}{10pt plus 3pt minus 3pt}{1pt plus 1pt minus 1pt}
\titlespacing\subsubsection{0pt}{8pt plus 3pt minus 3pt}{1pt plus 1pt minus 1pt}

\titleformat{\section}{\normalfont\large\bfseries}{\thesection}{1em}{}

\titleformat{\subsection}{\normalfont\normalsize\bfseries}{\thesubsection}{1em}{}

\titleformat{\subsubsection}{\normalfont\normalsize}{\thesubsubsection}{1em}{}

\titleformat{\paragraph}[runin]{\normalfont\normalsize\itshape}{\theparagraph}{1em}{}

\usepackage{mathtools,upgreek}
\usepackage{newtxtext,newtxmath} % use Times fonts if available on your TeX system

\usepackage[utf8]{inputenc}	% allow utf-8 input
\usepackage[T1]{fontenc}	% use 8-bit T1 fonts
\usepackage{xcolor}		% colors for hyperlinks
\usepackage[colorlinks = false]{hyperref}	% Color links to references, figures, etc.
\usepackage{booktabs} 		% professional-quality tables
\usepackage{nicefrac}		% compact symbols for 1/2, etc.
\usepackage{microtype}		% microtypography
\usepackage{lineno}		%   Line numbers
\usepackage{float}			% Allows for figures within multicol

\usepackage{siunitx}[=v2]
\usepackage{nicefrac,multirow}

\usepackage{subcaption}
\usepackage{algorithmic,algorithm2e}

\usepackage{enumitem}
\setlist[enumerate]{label*=\arabic*.} %sub-numbers 1.1 etc. in nested lists

\newcommand\numberthis{\addtocounter{equation}{1}\tag{\theequation}}
\mathtoolsset{centercolon}

\DeclareMathOperator{\sym}{sym}

\newcommand{\inte}[3]{\int \limits_{ #1} #2 \; \mathrm{d} #3} 
\newcommand{\binte}[4]{\mathop{\int}_{ #1}^{#2} #3 \; \mathrm{d} #4}

\newcommand{\ddp}{\, \partial} 

\newcommand{\nabr}{\nabla_{\ve{X}}}

\newcommand{\diffp}[2]{\frac{\partial #1}{\partial #2}}
\newcommand{\diffv}[2]{\frac{\delta #1}{\delta #2}}

\newcommand{\ve}[1]{\boldsymbol{#1}} 
\newcommand{\te}[1]{\mathbf #1}
\newcommand{\tg}[1]{\boldsymbol{#1}} 
\newcommand{\tte}[1]{\mathbb{#1}}

\newcommand{\tei}{\mathbf I}

\newcommand{\rset}{\mathbb{R}}
\newcommand{\nset}{\mathbb{N}}

\newcommand{\const}{\text{const}.}
\newcommand{\nv}{\mathrm}
\newcommand{\comma}{\hspace{3mm} \text{,}}

\newcommand{\glmand}{\hspace{3mm} \text{and} \hspace{3mm}}
\newcommand{\glmwith}{\hspace{3mm} \text{with} \hspace{3mm}}
\newcommand{\commam}{\hspace{3mm} \text{,} \hspace{3mm}}
\newcommand{\point}{\hspace{3mm} \text{.}}

\newcommand{\gc}{\mathcal{G}_\text{c}}
\newcommand{\gco}{\mathcal{G}_\text{c}^1}
\newcommand{\gct}{\mathcal{G}_\text{c}^2}
\newcommand{\rref}{r_\text{ref}}

\newcommand{\lc}{{\ell_\text{c}}}
\newcommand{\lcs}{{\ell^2_\text{c}}}
\newcommand{\etaf}{{\eta_\text{f}}}
\newcommand{\hist}{\mathcal{H}}

\newcommand{\bvi}{\beta_\text{vi}}

\newcommand{\Pilcd}{\dot \varPi_\lc}
\newcommand{\Pilcextd}{\dot \varPi^\nv{ext}_\lc}

\newcommand{\Pistlc}{\varPi^\nv{sd}_\lc}

\newcommand{\PiExtlc}{\varPi^\nv{ext}_\lc}

\newcommand{\iDvi}{{\mathcal{D}}^\text{vi}}
\newcommand{\iDfr}{{\mathcal{D}}^\text{fr}}

\newcommand{\Phifd}{\varPhi^\nv{fr}}
\newcommand{\Phifvd}{\varPhi^\nv{fr,vi}}
\newcommand{\Phivd}{\varPhi^\nv{vi}}
\newcommand{\PhivdX}[1][\varXi]{\prescript{}{#1}{\varPhi^\nv{vi}}}

\newcommand{\disd}{\dot{D}}
\newcommand{\dpfd}{\dot{D}^\nv{fr}}
\newcommand{\dvid}{\dot{D}^\nv{vi}}
\newcommand{\dvidX}[1][\varXi]{\prescript{}{#1}{\dot{D}^\nv{vi}}}

\newcommand{\psistoX}[1][\varXi]{\prescript{}{#1}{\psi^\text{st,ov}}}

\newcommand{\psistovX}[1][\varXi]{\prescript{\nv{vol}}{#1}{\psi^\text{st,ov}}}
\newcommand{\psistoiX}[1][\varXi]{\prescript{\nv{iso}}{#1}{\psi^\text{st,ov}}}

\newcommand{\kovX}[1][\varXi]{\prescript{}{#1}{\kappa^\nv{ov}}}

\newcommand{\NovX}[1][\varXi]{\prescript{}{#1}{N_\nv{O}^\nv{ov}}}

\newcommand{\muovX}[2][p]{\prescript{}{#2}{\mu^\nv{ov}_{#1}}}

\newcommand{\muovnX}[1][\varXi]{\prescript{}{#1}{\mu^\nv{ov}}}

\newcommand{\alovX}[2][p]{\prescript{}{#2}{\alpha^\nv{ov}_{#1}}}

\newcommand{\nuovnX}[1][\varXi]{\prescript{}{#1}{\nu^\nv{ov}}}

\newcommand{\viscX}[1][\varXi]{\prescript{}{#1}{\tte V}}

\newcommand{\etavX}[1][\varXi]{\prescript{\nv{vol}}{#1}{\eta}}
\newcommand{\etaiX}[1][\varXi]{\prescript{\nv{iso}}{#1}{\eta}}

\newcommand{\tauX}[1][\varXi]{\prescript{}{#1}{\tau}}

\newcommand{\felX}[1][\varXi]{\prescript{}{#1}{\te{F}^\nv{el}}}

\newcommand{\jelX}[1][\varXi]{\prescript{}{#1}{J^\nv{el}}}

\newcommand{\feliX}[1][\varXi]{\prescript{}{#1}{\overline{\te F}^\nv{el}}}

\newcommand{\fviX}[1][\varXi]{\prescript{}{#1}{\te{F}^\nv{vi}}}

\newcommand{\fvidX}[1][\varXi]{\prescript{}{#1}{\dot{\te{F}}^\nv{vi}}}

\newcommand{\belX}[1][\varXi]{\prescript{}{#1}{\te b}^\nv{el}}
\newcommand{\beldX}[1][\varXi]{\prescript{}{#1}{\dot{\te b}}^\nv{el}}

\newcommand{\beltrnX}[1][\varXi]{\prescript{}{#1}{{\te b}^\nv{el}_{n, \nv{tr}}}}
\newcommand{\belnX}[1][\varXi]{\prescript{}{#1}{{\te b}^\nv{el}_{n}}}

\newcommand{\celX}[1][\varXi]{\prescript{}{#1}{\tilde{\te C}}^\nv{el}}

\newcommand{\cviX}[1][\varXi]{\prescript{}{#1}{{\te C}^\nv{vi}}}
\newcommand{\cviXold}[1][\varXi]{\prescript{}{#1}{\te C_{n-1}^\nv{vi}}}

\newcommand{\lameX}[2][\beta]{\prescript{}{#2}{\lambda_{#1}^\nv{el}}}
\newcommand{\lamedX}[2][\beta]{\prescript{}{#2}{\bar{\lambda}_{#1}^\nv{el}}}
\newcommand{\lameiX}[2][\gamma]{\prescript{}{#2}{\bar \lambda^\nv{el}_{#1}}}

\newcommand{\lameqX}[2][\beta]{\prescript{}{#2}{{\lambda_{#1}^\nv{el}}^2}}
\newcommand{\nlameX}[1][\varXi]{\prescript{}{#1}{N^\nv{el}_\lambda}}
\newcommand{\nulameX}[2][\gamma]{\prescript{}{#2}{\nu}^\nv{el}_{#1}}

\newcommand{\pelX}[2][\beta]{\prescript{}{#2}{\te p_{#1}^\nv{el}}}
\newcommand{\PelX}[2][\beta]{\prescript{}{#2}{\tilde{\te P}_{#1}^\nv{el}}}

\newcommand{\dvieX}[1][\varXi]{\prescript{}{#1}{{\te d}^\nv{vi}}}

\newcommand{\lvitX}[1][\varXi]{\prescript{}{#1}{\tilde{\te{{l}}}{}^\nv{vi}}}
\newcommand{\dvitX}[1][\varXi]{\prescript{}{#1}{{\tilde {\te d}^\nv{vi}}}}

\newcommand{\lie}{\mathcal{L}}

\newcommand{\utaoX}[1][\varXi]{\prescript{0}{#1}{\tg \uptau}^\nv{ov}}

\newcommand{\uPoX}[1][\varXi]{\prescript{0}{#1}{\te P}^\nv{ov}}
\newcommand{\uP}[1][\varXi]{\prescript{0}{}{\te P}}

\newcommand{\wu}{\delta \ve u}

\newcommand{\wus}{\mathbb{W}_{\ve u}}

\newcommand{\wc}{\delta c}

\newcommand{\wcs}{\mathbb{W}_c}

\newcommand{\omref}{\varOmega_{0}}
\newcommand{\domref}{\partial \varOmega_{0}}
\newcommand{\domrefu}{\partial \varOmega_{0,\hat{\ve u}}}
\newcommand{\domrefp}{\partial \varOmega_{0,\hat{\ve p}}}

\newcommand{\veZero}{\textbf{\textit{0}}}

\usepackage{multicol}

\newcommand{\urbd}{{\dot u}} %displacement rate prescribed in SENT, DENT

\DeclareMathOperator*{\stat}{stat}

\usepackage{mdframed}

\title{Rate- and temperature-dependent ductile-to-brittle fracture transition: Experimental investigation and phase-field analysis for toffee}

\begin{document}

\author[a]{Franz Dammaß}
\author[b]{Dennis Schab}
\author[b]{Harald Rohm}
\author[,a]{Markus Kästner \thanks{Contact: \texttt{markus.kaestner@tu-dresden.de} }}
\affil[a]{Institute of Solid Mechanics, TU Dresden, Germany}
\affil[b]{Chair of Food Engineering, Institute of Natural Materials Technology, TU Dresden, Germany} 
\date{}
\maketitle

\begin{abstract}

The mechanical behaviour of many materials, including polymers or natural materials, significantly depends on the rate of deformation.
As a consequence, a rate-dependent ductile-to-brittle fracture transition may be observed.
For toffee-like caramel, this effect is particularly pronounced. At room temperature, this confectionery may be extensively deformed at low strain rates, while it can behave highly brittle when the rate of deformation is raised.
Likewise, the material behaviour does significantly depend on temperature, and even a slight cooling may cause a significant embrittlement. 

In this work, a thorough experimental investigation of the rate-dependent deformation and fracture behaviour is presented.
In addition, the influence of temperature on the material response is studied.
The experimental results form the basis for a phase-field modelling of fracture. In order to derive the governing equations of the model, an incremental variational principle is introduced.
By means of the validated model, an analysis of the experimentally observed ductile-to-brittle fracture transition is performed.
In particular, the coupling between the highly dissipative deformation behaviour of the bulk material and the rate-dependent fracture resistance is discussed.

\vspace{5mm}
\noindent
\textbf{Keywords: } Fracture~\textendash~Rate-dependent Ductile-to-brittle transition~\textendash~Caramel~\textendash~Toffee~\textendash~Phase-field~\textendash~Incremental variational principle

\end{abstract}

\section{Introduction}
\label{sec:intro}

\begin{figure}[p!]
\begin{mdframed}
\begin{center}
\textbf{Nomenclature}
\end{center}

Within this paper, italic symbols are used for scalar quantities ($d$, $\varPsi$) and bold italic symbols for vectors ($\ve u$). For Second-order tensors, bold non-italic letters ($\te T$, $\tg \uptau$) are used, whereas fourth-order tensors are written in \textit{Blackboard bold}~($\tte C$).

\begin{multicols}{2}
\small
% 
%     \begin{center}
%     \ \\
%     \begin{minipage}[b]{0.48\textwidth}
    \begin{tabular}{l p{0.35\textwidth}}
        \multicolumn{2}{l}{Latin symbols} \\
        $\te b$ & left Cauchy-Green deformation tensor \\
        $c$ & parameter of the rate-dependent fracture toughness function\\
        $\te C$ & right Cauchy-Green deformation tensor \\
        $d$ & fracture phase-field variable \\
        $\te d$ & rate of deformation\\
        $\dot D$ & density of dissipation power\\
        $\te F$ & deformation gradient\\
        $g$ & degradation function \\
        $\gc$ & fracture toughness \\
        $\gco$ & parameter of the rate-dependent fracture toughness function\\
        $\gct$ & parameter of the rate-dependent fracture toughness function\\
        $\hist$ & history variable \\
        $\tte H^1$ & Sobolev space \\
        $\te I$ & second-order identity tensor\\
        $\tte I^\text{D}$ &  symmetric fourth-order deviator projection tensor\\
        $J$ & determinant of the deformation gradient\\
        $k$ & residual stiffness \\
        $\lc$ & characteristic width of diffuse crack topology\\
        $\te l$ & velocity gradient\\        
        $N$ & number of dimensions in space \\
        $\nlameX$ & number of pair-wise different elastic principal stretches \\
        $\NovX$ & number of Ogden terms\\
        $\ve N$ & outward-pointing normal \\
        $\mathbb N$ & set of natural numbers \\
        $p$ & Ogden term index \\
        $\ve p$ & Piola traction vector \\
        $\te P$ & first Piola-Kirchhoff stress tensor \\
        $\te p_\beta$ & current eigenvalue basis tensor belonging to $\lambda_\beta$\\
        $\te P_\beta$ & reference eigenvalue basis tensor belonging to $\lambda_\beta$\\
        $r$ & scalar equivalent rate of deformation \\
        $\rref $ & parameter of the rate-dependent fracture toughness function\\
        $t$ & time  \\
        $\te T$ & second Piola-Kirchhoff stress tensor \\
        $\ve u$ & displacement \\
        $\tte V$ & viscosity tensor \\
        $\tte W$ & test function space \\
        $\ve x$ & current Cartesian coordinate \\
        $\ve X$ & reference Cartesian coordinate\\
                & \\
        \multicolumn{2}{l}{Operators} \\
        $\dot \circ$ & time-derivative of $\circ$\\
        $\partial \circ$ & boundary of $\circ$\\
        $\nabr $ & Nabla operator with respect to $\ve X$ \\
        $\lie \, \circ$ & Lie derivative of $\circ$\\
        $\sym \circ$ & symmetric part of $\circ$ \\
    \end{tabular}
%     \end{minipage}%
    
%     \begin{minipage}[b]{0.48\textwidth}
    \begin{tabular}{l p{0.35\textwidth}}
        \multicolumn{2}{l}{Greek symbols} \\
        $\alpha$ & Ogden model exponent\\
        $\gamma_{\lc}$ & crack surface density\\
        $\delta$ & Kronecker delta\\
        $\wu$ & test function for the mechanical equilibrium \\ 
        $\wc$ & test function for the phase-field equation\\
        $\etaf$ & artificial fracture viscosity \\
        $\eta$ & viscosity \\
        $\vartheta$ & temperature\\
        $\kappa$ & bulk modulus \\
        $\lambda$ & principal stretch \\
        $\mu$ & shear modulus\\
        $\nu$ & {Poisson's} ratio\\
        $\varXi$ & over-stress branch index\\
        $\dot \varPi_{\lc}$ & rate type pseudo-potential \\
        $\tau $ & relaxation time \\
        $\tg \uptau$ & Kirchhoff stress tensor\\
        $\varPhi$ & dissipation potential\\
        $\ve \chi$ & motion function\\
        $\psi$ & density of virtually undamaged free energy\\
        $\varPsi$ & density of free energy \\
        $\varOmega$ & domain\\ 
        
    & \\
    \multicolumn{2}{l}{Sub-/superscripts} \\
        $\circ_0$ & quantity $\circ$ refers to the reference configuration\\
        ${}^0\circ$ & virtually undamaged value of quantity $\circ$ \\
        $\tilde \circ$ & quantity $\circ$ refers to the inelastic intermediate configuration \\
        $\bar \circ$ & isochoric portion of $\circ$\\
        $\hat \circ$ & prescribed value of quantity $\circ$\\
        el & elastic\\
        ext & external forces\\
        fr & fracture\\
        fr,vi & viscous regularisation of crack growth\\
        iso & isochoric\\
        ov & over-stress \\
        tr & trial \\
        vi & viscous \\
        vol & volumetric\\

        & \\
        \multicolumn{2}{l}{Abbreviations} \\
        CZE & Cohesive zone element \\
        DENT & Double edge-notched tension test\\
        DIC & Digital image correlation\\
        DMA & Dynamic mechanical analysis \\
        FE & Finite element \\
        SENB & Single edge-notched bending test \\
        SENT & Single edge-notched tension test \\
        XFEM & Extended Finite element method \\
    \end{tabular}
%     \end{minipage}

%     \end{center}

\end{multicols}
\end{mdframed}
\end{figure}

For optimising processes in food production and packaging, a predictive model that adequately describes the mechanical behaviour of the respective material can essentially facilitate procedures, which are often realised in purely empirical manner, so far.
For instance, it may be a non-trivial question to determine the optimal cutting speed for foods or candies such as toffee within an industrial process. Whereas slow cutting is economically undesirable, choosing a too high speed might result in rough surfaces or even splintering, cf.~\cite{schuldt2018,schab2021}.
Compared to many classical engineering materials, the chemical composition and the microstructure of toffee-like confections are highly complex.%
\footnote{There are different notions of the term \textit{toffee} in different parts the world, see \cite[Chapter~10]{hartel2018}. In this work, a candy that is chewy at room temperature is considered, and both the denominations \textit{toffee} and \textit{caramel} are used.
Its moisture content is of \mbox{(7.0  g ± 0.2 g)/100 g} and the precise composition is given in Sect.~\ref{sec:mat}.}
Basically, these caramels consist of a sugar matrix, a protein network, water and fat \cite{schab2021,hartel2018}, and the mechanical properties result from the interplay of these constituents.
Accordingly, some analogies can be drawn to filled polymers as considered for technical applications, cf.~\cite{schuldt2018}, but, due to the significant content of moisture and fat, also to fluids.
In addition, some micro- or mesoscopic heterogeneity and a relevant amount of scatter of the mechanical properties are inherent to these natural materials.

In the food engineering community, several publications address the rate-dependency of deformation of natural materials and the determination of fracture mechanics properties of foods, see e.g.~\cite{vanvliet2013,schuldt2018} for an overview.
Experimental studies have revealed the significance of viscous effects when typical food-like materials are deformed, for instance toffee, cheese, butter, salami or bread dough \cite{rohm1992,rohm1993,vanvliet2013,schuldt2016}.
Most commonly, these phenomena are investigated by means of \textit{Dynamic mechanical analysis (DMA)}. Typically, these experiments are performed under either uniaxial tension, compression or torsion with small sinusoidal deformations prescribed at varied frequency, cf.~\cite{menard2020}. Accordingly, numerous results are available in terms of rate-dependent viscoelastic moduli. Likewise, in the scope of \textit{DMA}, the influence of temperature is studied by means of temperature variations, and analogies between frequency and temperature dependency of the mechanical behaviour, for instance for sugar-based confections such as toffee \cite{schmidt2018}.
However, so far, less efforts have been devoted to the generalisation of these \textit{DMA} results to three-dimensional constitutive models. Moreover, the clear separation between different dissipative mechanisms such as viscous effects, plasticity, contact friction and damage or fracture remains a challenging issue.
Due to the complexity of failure in foods and other dissipative soft solids, the determination of their fracture mechanical properties is a subject of ongoing research.
Specific setups are proposed in order to characterise the resistance against fracture, see \cite{goh2002} for an overview. For instance, methods that are based on cutting of the material using wires of varied diameters~\cite{kamyab1998,goh2005,gamonpilas2009,forte2015} or blades~\cite{patel2009c,skamniotis2016a} are employed. Alternative approaches include the theory of \textit{Essential work of fracture}~\cite{williams2007a,skamniotis2017},  and the analysis of pre-notched specimens under tension \textit{(SENT)} or bending \textit{(SENB)}~\cite{kamyab1998,gamonpilas2009}.
For several foods such as different types of cheese~\cite{goh2005} or starch gels~\cite{luyten1995,gamonpilas2009} and gelatine gels~\cite{forte2015}, experimental results indicate rate-dependency of the fracture resistance.
More recently, numerical simulations of damage and fracture processes in food have also been presented. For instance, cutting processes have been modelled employing \textit{Cohesive zone elements}~\textit{(CZE)}~\cite{goh2002,goh2005,boisly2016} or  node-release techniques together with strain criteria~\cite{gamonpilas2009}, and continuum-damage models have been used for the simulation of mastication using commercial finite element software~\cite{skamniotis2017a,samaras2023}.
While these numerical investigations have been performed for cheese~\cite{goh2002,goh2005} and cheese-like model materials~\cite{boisly2016}, and starch-based foods~\cite{gamonpilas2009,skamniotis2017a,samaras2023}, for instance, no models for toffee-like caramels are available, though. In particular, there is a lack for an adequate model of the rate-dependent ductile-to-brittle fracture transition as it can be observed in experiments for this specific class of natural materials.

In the mechanics community, the phase-field approach has become a well-established concept for the modelling of fracture phenomena. Different from classical finite element approaches \textit{(FE)}, it enables to simulate crack growth without the need for remeshing and crack patterns that are not a priori known can be simulated in a straightforward manner, which makes the concept attractive compared to alternative approaches such as \textit{CZE}~\cite{ortiz1999} or the \textit{Extended-finite-element-method} \textit{(XFEM)} \cite{moes2017}.
The phase-field fracture approach goes back to the variational formulation of brittle fracture \cite{francfort1998}, which is based on the Griffith criterion \cite{griffith1921} for crack propagation.
Introducing a diffuse approximation of crack topology, a regularisation of the underlying pseudo-energy functional can be carried out~\cite{bourdin2000,bourdin2008}. Descriptively, a \textit{smeared} crack representation by means of the phase-field variable is adopted, which continuously varies from the intact to the fully broken material state and cracks are no longer seen as sharp discontinuities, but approximated over a finite length scale $\lc$.
Linear elasticity and small deformations are assumed for the original regularised variational formulation \cite{bourdin2000}, which $\Gamma$-converges towards classical Linear-elastic fracture mechanics for $\lc \rightarrow 0$.
Nevertheless, numerous modified and extended phase-field models of fracture have been developed, that do not necessarily provide $\Gamma$-convergence and variational consistency, yet enable to account for more complexity in the constitutive behaviour.
For the modelling of brittle fracture, several advancements are available, which, for instance, aim at accounting for crack closure effects \cite{miehe2010a,amor2009,steinke2019}, anisotropy of the fracture resistance  \cite{clayton2014a,teichtmeister2017,nguyen2017a} or failure-mode dependency of the toughness \cite{zhang2017,hug2022}, and interfacial failure \cite{hansen-doerr2020,schoeller2022}, naming just a few.
Moreover, different extensions of the initial theory have been suggested in order to account for fatigue effects, e.g. \cite{carrara2020,seiler2020,schreiber2020}, see~\cite{kalina2023a,li2023} for an overview.
Furthermore, various phase-field models of fracture in elasto-plastic materials, have been proposed with different approaches for the coupling between the dissipative mechanisms in the bulk and fracture, including \cite{miehe2015,kuhn2016,miehe2016a,ambati2016,borden2016,yin2020c}. A comprehensive comparison of several approaches is provided in \cite{alessi2018a} and, recently, \cite{marengo2023}.

The phase-field modelling of rate-dependent fracture phenomena also is a subject of current research and an overview is provided in the authors' previous contributions \cite{dammass2021b,dammass2023}.
Several models aiming at the description of failure of materials with rate-dependent bulk deformation behaviour have been presented within the kinematically linear regime \cite{liu2018,shen2019}, the framework of linear viscoelasticity at finite deformation \cite{loew2019,loew2020}, and finite viscoelasticity \cite{yin2020,brighenti2021,arash2021}.
Thereby, different approaches for the coupling between inelastic mechanisms and damage have been pursued, which have been studied in \cite{dammass2021b,dammass2023} by means of a unified model. As a result, in \cite{dammass2023}, it has been shown that a fracture driving force contribution related to accumulated viscous dissipation can lead to model predictions that deviate from experimental evidence.
Compared to the rate-dependency of deformation of the bulk, less efforts have been devoted to the modelling of strain rate-dependent resistance against fracture.
In order to investigate the brittle-to-ductile fracture transition observed in metals under shear loading, a phenomenological ansatz for the rate-dependent toughness of metals has been introduced in \cite{miehe2015}, where the bulk deformation behaviour is assumed to be rate-independent.
Likewise, in \cite{yin2020a}, the toughness of a linear elastic material is assumed to depend on rate of deformation.
Moreover, in the recent contribution \cite{dammass2023}, a phase-field model that combines both a rate-dependent, viscoelastic bulk resistance and a rate-dependent fracture resistance has been investigated and its suitability for simulating rate-dependent transitions between ductile and brittle behaviour has been shown by means of numerical examples.

The present contribution aims at the investigation of the ductile-to-brittle fracture transition as it can be observed for toffee-like caramel.
To this end, an extensive experimental study on the rate-dependency of the mechanical behaviour is performed, for which a toffee type caramel confectionery is produced under lab conditions.
Furthermore, the influence of temperature on the material response is studied.
For the description and model-based analysis of the rate-dependent fracture phenomena, the phase-field approach is pursued. In order to introduce the governing equations of the model, an incremental variational principle is presented, which is based on a time-continuous pseudo-potential of rate type, cf.~\cite{miehe2011}.
As a result, model equations are obtained that correspond to a specification of the unified model presented in the previous work \cite{dammass2023}.

This contribution proceeds as follows. In Sect. \ref{sec:model}, the key assumptions of the model are presented and governing equations are derived.
For this purpose, an incremental variational principle is introduced.
Furthermore, thermodynamic consistency is proven and key aspects of the finite element implementation are addressed.
Subsequently, in Sect.~\ref{sec:mat}, information on the toffee-like material that is considered in the scope of this paper is presented.
Then, in Sect.~\ref{sec:mech-25}, an analysis of its deformation and fracture behaviour at room temperature is performed, and subsequently, in Sect.~\ref{sec:mech-18}, the influence of temperature on the mechanical behaviour is studied.
The paper closes with a conclusion and an outlook regarding future works in Sect.~\ref{sec:concl}. 

\section{Phase-field model of rate-dependent fracture at finite deformation}
\label{sec:model}

In the previous works of the authors \cite{dammass2021a,dammass2021b}, the coupling between a viscoelastic bulk material and the fracture phase-field has been studied in the kinematically linear regime by means of a unified model.
Moreover, in \cite{dammass2023}, the work has been extended to the framework of finite viscoelasticity \cite{reese1998}, and, in addition to the viscoelastic model of bulk deformation, a fracture resistance that depends on the rate of deformation has been considered in a numerical study.

In the following, one specification of the unified approach considered in the authors' previous contributions will be applied to  the toffee-like confectionery.
Extending the previous work, the model is formulated in the framework of an incremental variational principle, which serves for the derivation of the governing differential equations.

\subsection{Diffuse crack representation and fracture dissipation}

\begin{figure}[tbp!]
		\centering 
		\includegraphics[width=0.8\linewidth,trim=0mm 2mm 0mm 5mm]{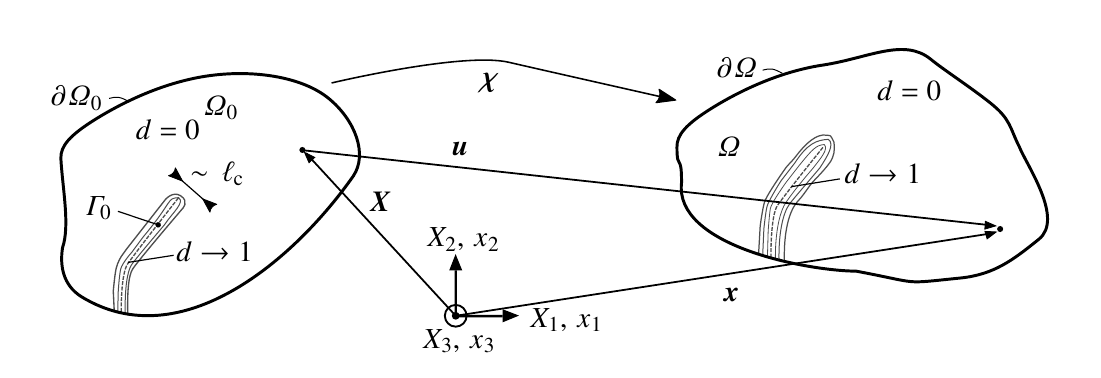} % [trim=left bottom right top]
		\caption{Diffuse representation of a crack within a domain that undergoes finite deformation}
		\label{fig:kinem}
\end{figure}

The starting point for the present model of rate-dependent fracture phenomena is the diffuse approximation of crack topology by means of the fracture phase-field variable
\begin{equation}
 d: \omref \times [0,t] \rightarrow [0,1] \comma \hspace{3mm} (\ve{X}, t) \mapsto d(\ve{X},t)
 \label{eq:d-def}
\end{equation}
that continuously varies from the intact ($d=0$) to the fully broken ($d=1$) material state, wherein \mbox{$\ve X \in \omref$} denotes the coordinate of a material point in the reference configuration and $t$ the time.
This concept, which goes back to the fundamental work of Bourdin \cite{bourdin2000}, is illustrated by Fig.~\ref{fig:kinem}.
Using the phase-field variable, following Miehe et al.~\cite{miehe2010},
a \textit{crack surface density} functional
\begin{equation}
 \gamma_\lc = \frac{1}{4 \, \lc} \left( d^2 + 4 \, \lcs \, \nabr d \cdot \nabr d \right)
 \label{eq:surfDen}
\end{equation} 
can be defined, in which the length scale parameter~$\lc$ controls the characteristic width of the diffuse crack. Therein, the Nabla operator with respect to the reference coordinate $\ve X$ is given by
\begin{equation}
 \nabr \, \circ = \sum_{K=1}^N \ve e_K \, \left(\diffp{\, \circ}{X_K}\right) \comma
\end{equation}
with $\ve e_K$ denoting the $K$-th basis vector of the Cartesian reference coordinate frame.
Based on this functional and conceptionally similar to the hypothesis of Griffith~\cite{griffith1921}, the density of rate of dissipation due to crack evolution is defined as
\begin{equation}
 \Phifd = \dot \gamma_\lc \, \gc \comma
 \label{eq:Phifd-def}
\end{equation}
which refers to a volume element of the reference or undeformed configuration  $\omref \subset \rset^N$ of the $N$-dimensional domain under consideration. In $\Phifd$, the parameter $\gc>0$ quantifies the resistance of the material against fracture.
In analogy to the proportionality constant of classical fracture mechanics, $\gc$ is referred to as fracture toughness in this contribution. This is in line with the authors' previous works \cite{dammass2021b,dammass2023} and other contributions in the literature, see e.g. \cite{miehe2015b,yin2020c,yin2020a,hansen-dorr2020}.%
\footnote{It has to be noted that the proposed model does not guarantee $\Gamma$-convergence against a \textit{sharp} crack description. Therefore, $\gc$ can be seen as a measure of resistance against fracture that is valid within the diffuse framework, yet can not be assumed to exactly correspond to the toughness or critical energy release rate according to classical fracture mechanics, in general.}
While $\gc$ classically is assumed to be a constant, it may also depend on the state of deformation or deformation history, respectively. In particular, $\gc$ can depend on the rate of deformation, cf. \cite{dammass2023,miehe2015,yin2020a}. Moreover, it may also change during fatigue life, see e.g. \cite{seiler2020}, or due to plastic deformation \cite{yin2020c,han2021}. 
In the following, the phase-field framework is set up for $\gc=\const$, first. Subsequently, the model is extended to account for a fracture toughness that depends on rate of deformation in Sect.~\ref{sec:gov_eqs}.

In order to enhance the stability of the numerical solution procedure, a viscous regularisation of crack growth is assumed. Therefore, in addition to the dissipation due to the increase in crack surface, which is accounted for by means of $\Phifd$, an additional contribution to the fracture dissipation
\begin{equation}
 \Phifvd = \frac{1}{2} \, \etaf \, {\dot d}^2
\end{equation}
is defined, cf. \cite{kuhn2013}. The fracture viscosity $\etaf >0$ is chosen such small that its influence on the predicted responses vanishes, which is verified by means of a numerical study.

The decrease of free energy due to fracture is expressed by means of the degradation function
\begin{equation}
 g: [0,1] \rightarrow [0,1] \hspace{10mm} \comma \hspace{10mm} d \mapsto (1-k) \, (1-d)^2 + k
 \comma
 \label{eq:degfu-def}
\end{equation} 
cf. \cite{bourdin2000,miehe2010,kuhn2016,shen2019}, which fulfils the conditions
\begin{align*}
 g(d=0)=1 \commam g(d=1)=0 \commam  \frac{\ddp g}{\ddp d} \leq 0 \commam \left.\frac{\ddp g}{\ddp d}\right|_{d=1}=0 \comma
 \numberthis
 \label{eq:degfu-bed}
\end{align*}
and in which a small residual $k$ is included in order to enhance numerical stability.
By means of $g(d)$, the degraded reference free energy density can be written as
\begin{equation}
 \varPsi = g(d) \, \psi \comma
\end{equation} 
with the \textit{virtually undamaged} density of free energy, i.e. the amount of free energy that would be stored in the material in the absence of damage, denoted by $\psi$.
\subsection{Rate-dependent bulk response}
\begin{figure}[b!]
		\centering 
		\includegraphics[scale=0.9,trim=0mm 3mm 5mm 0mm]{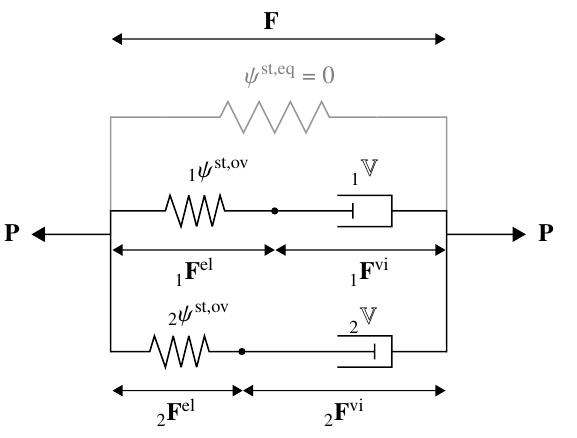} % [trim=left bottom right top]
		\caption{Generalised Maxwell element\textemdash Constitutive assumptions}
		\label{fig:maxwell}
\end{figure}
For the description of the rate-dependent deformation behaviour, the approach of finite viscoelasticity in the sense of \cite{reese1998} is pursued.
Accordingly, material behaviour can be illustrated by means of a Generalised Maxwell element as presented in Fig.~\ref{fig:maxwell}. Here, the special case of zero equilibrium stiffness is considered, which is discussed with reference to experimental data in Sects.~\ref{sec:mech-25} and~\ref{sec:mech-18}.

\subsubsection{Kinematics}
The displacement of a material point is denoted by
\begin{equation}
 \ve u(\ve X,t) = \ve \chi(\ve X,t) - \ve X \comma
\end{equation}
wherein
\begin{equation}
 \ve \chi(\ve X,t) : \omref \times [0,t] \rightarrow \varOmega \comma \hspace{3mm} (\ve X, t) \mapsto \ve x(\ve X,t)
\end{equation} 
is the motion function, which can be assumed to be bijective and continuous in space and time.
The deformation gradient $\te F$ and its determinant $J$ are then given by
\begin{equation}
 \te F = \left(\nabr \ve \chi\right)^\top \qquad \text{and} \qquad J= \det \, \te F > 0 \comma
\end{equation}
and the rate of deformation tensor is given by
\begin{equation}
 \te d = \sym \left( \dot{\te F} \cdot \te{F}^{-1} \right) \point
\end{equation}
For the description of toffee, a Generalised Maxwell element that consists of two non-equilibrium branches has shown to be suitable, see Sects.~\ref{sec:mech-25} and~\ref{sec:mech-18}. For each of these over-stress branches, $\varXi \in \{1, 2\}$, the deformation gradient is multiplicatively decomposed into an elastic and a viscous part, i.e.%
\footnote{It has to be noted that the Einstein summation convention does not apply for the index of the non-equilibrium branch $\varXi$ in \eqref{eq:inel-split} and the following kinematic relations.}
\begin{equation}
\te F = \felX \cdot \fviX \point
\label{eq:inel-split}
\end{equation}

Following Flory~\cite{flory1961}, for these tensors of each non-equilibrium branch, portions can be defined that account for the corresponding isochoric part of deformation, e.g.
\begin{equation}
 \feliX =  {\jelX}^{-1/3} \, \felX \qquad \text{with} \qquad \jelX = \det \, \felX \point
  \label{eq:floryEl}
\end{equation} 
Furthermore, for each of the non-equilibrium branches, the elastic stretch tensors 
\begin{equation}
 \belX = \felX \cdot {\felX}^\top 
 \qquad \text{and} \qquad
 \celX = {\felX}^\top \cdot {\felX}
  \label{eq:stretchEl}
\end{equation} 
are defined, where the tilde symbol $\tilde \circ$ is introduced to mark quantities which refer to a fictitious inelastic intermediate configuration defined by $\fviX$.
From the spectral decomposition of these quantities,
\begin{equation}
 \belX = \sum\limits_{\beta=1}^{\nlameX} \lameqX[\beta]{\varXi} \, \pelX[\beta]{\varXi}
 \qquad \text{or} \qquad
 \celX = \sum\limits_{\beta=1}^{\nlameX} \lameqX[\beta]{\varXi} \, \PelX[\beta]{\varXi}
 \comma
\end{equation} 
the elastic principal stretches $\lameX[\beta]{\varXi}$ as well as their isochoric counterparts $\lamedX[\beta]{\varXi} = {\jelX}^{-1/3} \, \lameX[\beta]{\varXi}$ can be identified, where the number of pair-wise different elastic principal stretches is given by $\nlameX$ for the corresponding non-equilibrium branch.
The according second-order projection tensors, or \textit{eigenvalue-base} tensors, are obtained from
\begin{equation}
\pelX[\beta]{\varXi} = \delta_{1 \nlameX} \, \, \te I \, + \prod\limits_{\gamma} \frac{\belX - \lameqX[\gamma]{\varXi} \, \tei}{\lameqX[\beta]{\varXi} - \lameqX[\gamma]{\varXi}}
\quad , \quad 
\gamma \in [1,\nlameX]\setminus \beta \subset \nset
\comma
 \label{eq:projtens}
\end{equation}
in which $\tei$ designates the second-order identity tensor, and the Kronecker delta $\delta_{kl}$ is given by
\begin{equation}
 \delta_{kl} = 
 \left\{ \begin{array}{c}
            1, \quad k = l \\
            0, \quad k \neq l \\ 
        \end{array}
 \right.
 \comma
 \label{eq:krondelta}
\end{equation}
and similar relations for $\PelX[\beta]{\varXi}$ \cite{miehe1993,miehe1998}.%

In order to describe rate of inelastic deformation within each non-equilibrium branch, the inelastic velocity gradient
\begin{equation}
 \lvitX = \fvidX \cdot {\fviX}^{-1}
 \label{eq:lvit-def}
\end{equation}
is introduced. It refers to the respective intermediate configuration defined by $\fviX$.
Here, it is assumed that the inelastic deformation is purely translational, i.e. the inelastic spin is defined to zero, which is a common definition for isotropic materials, see e.g. \cite{dafalias1984,wriggers2008,desouzaneto2008}.
Accordingly, the rate of viscous deformation can be written as
\begin{equation}
 \dvitX = \sym \lvitX = \lvitX \comma
 \label{eq:dvit-def}
\end{equation}
with its counterpart transformed to the reference configuration given by
\begin{equation}
 \dvieX = \felX \cdot \dvitX \cdot {\felX}^{-1} \point
 \label{eq:dvi-def}
\end{equation}
The rate of viscous deformation can also be written in terms of the Lie derivative of the elastic stretch $\lie\left[ \belX \right]$,
\begin{equation}
 \dvieX =  - \frac{1}{2} \,\lie\left[ \belX \right] \cdot {\belX}^{-1}
 \comma
 \label{eq:kinRel_lie_dvi}
\end{equation}
where the Lie derivative is given as
\begin{equation}
 \lie\left[ \belX \right] = \te F \cdot \left[ \left(\cviX\right)^{-1} \right]^{\boldsymbol{\cdot}} \cdot \te F^\top
   \quad \glmwith \quad
 \cviX = {\fviX}^\top \cdot \fviX
 \comma
\end{equation} 
cf.~\cite[Chapter 14]{desouzaneto2008} for a rigorous discussion in the context of elasto-plasticity.

\subsubsection{Strain energy and viscous dissipation}
The density of free energy $\varPsi$ is assumed to be given by the strain energy which can be associated to the springs of the two non-equilibrium branches of the rheological model shown in Fig.~\ref{fig:maxwell}.%
\footnote{It seems worth mentioning that in the previous work \cite{dammass2023}, depending on the choice of a scalar weighting parameter $\bvi$, an additional contribution to $\varPsi$ has been considered, which is related to the viscous dissipative mechanisms. As a consequence, a portion of accumulated viscous dissipation entered the fracture driving force. However, it also has been shown in \cite{dammass2023} that such a contribution can be disadvantageous when aiming at the prediction of experimental responses of rubbery polymers.
Likewise, although the deformation behaviour of toffee does significantly differ from an elastomer, a fracture driving force contribution related to accumulated viscous dissipation would not be suitable for toffee, cf. Sects.~\ref{sec:mech-25} and~\ref{sec:mech-18}.
Nevertheless, the present approach is retained in the unified model as presented in \cite{dammass2023}, for which the corresponding parameter has to be set to $\bvi=0$.}
Accordingly,
\begin{equation}
 \varPsi \left(\te F, \fviX, d \right) = g(d) \, \sum\limits_{\varXi=1}^{2} \psistoX \left( \te F, \fviX \right)
 = g(d) \, \psi
 \label{eq:Psi_ans}
\end{equation}
is defined for the density of free energy with respect to a volume element in the reference configuration.%
\footnote{As it is required from the principle of objectivity, $\psi$ and $\varPsi$, respectively, do not directly depend on $\te F$ and $\fviX$, cf. \cite{holzapfel2000}, yet on objective quantities that are computed from $\te F$ and $\fviX$ as specified in the concrete definition \eqref{eq:psisto}. Nevertheless, in order to simplify notation,  $\varPsi (\te F, \fviX, d )$ is written in the remainder of this contribution.}
For the \textit{virtually undamaged} strain energy of the individual over-stress branches, functions of compressible Ogden type \cite{ogden1997} are defined, i.e.
\begin{align*}
 \psistoX =& \psistovX +\psistoiX \\
 =& \, \frac{\kovX}{4} \left({\jelX}^2 - 2 \ln \jelX -1 \right)
 + \sum_{p=1}^{\NovX} \frac{\muovX[p]{\varXi}}{\alovX[p]{\varXi}} \, 
		\left(
		\sum_{\sigma=1}^{\nlameX} \nulameX[\sigma]{\varXi} \, \left(\lameiX[\sigma]{\varXi} \right)^{\alovX[p]{\varXi}} -3
        \right)
        \comma
        \numberthis
        \label{eq:psisto}
\end{align*}
wherein $\lameiX[\sigma]{\varXi} = \jelX^{-1/3} \, \lameX[\sigma]{\varXi}$ are the isochoric elastic principal stretches following from~\eqref{eq:floryEl} with their algebraic multiplicity given by $\nulameX[\sigma]{\varXi} \in \{1, 2, 3\}$.
Furthermore, for each over-stress branch, the compression moduli are denoted by  $\kovX >0$, and $\NovX$, $\alovX[p]{\varXi}$, $\muovX[p]{\varXi} >0$ are parameters of the Ogden models.
From these constants, the initial shear moduli and the according Poisson's ratios
\begin{equation}
 \muovnX = \frac{1}{2} \sum_{p=1}^{\NovX} {\muovX[p]{\varXi}} \, {\alovX[p]{\varXi}}
 \qquad \text{and} \qquad
 \nuovnX = \frac{3 \, \kovX - 2 \, \muovnX}{2 \, (3 \, \kovX + \muovnX)}
\end{equation}
can be derived.

The density of rate of viscous dissipation $\Phivd$ can be expressed by means of a quadratic form for each non-equilibrium branch
\begin{equation}
 \Phivd = \sum\limits_{\varXi=1}^{2} \, {\frac{1}{2} \, \dvieX \, : g(d) \viscX \, : \dvieX}
 = \sum\limits_{\varXi=1}^{2} \, \PhivdX \comma
 \label{eq:vis-dissP} 
\end{equation} 
in which $\viscX$ denote fourth-order positive definite viscosity tensors, cf. \cite{reese1998}. Note that the contributions of each over-stress branch $\PhivdX$, respectively, are strictly convex and take the minimum $\PhivdX(\dvieX = \te 0)$.
In the sense of Biot's fundamental work \cite{biot1965} and the \textit{concept of generalised standard materials} \cite{halphen1975}, $\Phivd$ or $\PhivdX$, respectively, correspond to typical dissipation potentials, which are related to the inelastic deformation, see also \cite{kumar2016}.
Here, isotropy is assumed and
\begin{equation}
 {\viscX}= 2 \, \etaiX \, \, \tte I^\text{D} + 9 \, \etavX \, \,  \te I \otimes \te I
\end{equation} 
is defined, accordingly. Therein, $\tte I^\text{D}$ is the symmetric fourth-order deviator projection tensor with its coordinates given by
\begin{equation}
\tte I^\text{D}_{klmn} = \frac{1}{2} \left( \delta_{km} \delta_{ln}+\delta_{kn} \delta_{lm} \right) - \frac{1}{3}\, \delta_{kl} \delta_{mn}
\point
\end{equation}
Moreover, $\etaiX, \, \etavX > 0$ are constant viscosities with respect to the isochoric and volumetric portion of deformation, respectively.
Here, these are identified from the relaxation times
\begin{equation}
 \tauX = \frac{\etaiX}{2 \, \muovnX} = \frac{\etavX}{\kovX} \comma
\end{equation} 
which are prescribed to be identical for isochoric and volumetric deformations.

\subsection{Pseudo-potential and governing equations}
\label{sec:gov_eqs}
Based on the previous assumptions on the densities of stored energy and dissipation power, a rate type pseudo-potential
\begin{equation}
\Pilcd = \inte{\omref}{ \left[ \dot \varPsi + \Phifd +\Phifvd + \Phivd \right]}{V} + \Pilcextd
\end{equation}
can be defined,%
\footnote{For the scope of the present study, inertia terms are neglected.}
in which the power of external forces is given by
\begin{equation}
 \Pilcextd = - \inte{\partial \, \omref}{\hat{\ve{p}} \cdot \dot{\ve u} }{A}
 \comma
\end{equation} 
with $\hat{\ve{p}}$ denoting the Piola traction vector prescribed on $\domrefp \subset \domref$.%
\footnote{In this work, no volume forces are considered. Furthermore, the external loads are assumed not to depend on the field variables.}
Furthermore, in $\Pilcd$, making use of the definition \eqref{eq:Psi_ans}, the rate of free energy can be written as
\begin{align*}
 \dot \varPsi   &= \diffp{\varPsi}{\te F} \, : \dot{\te F}^\top 
                    + \sum\limits_{\varXi=1}^{2} \, \diffp{\, \, \varPsi}{\fviX} \, : {\fvidX}^\top
                    + \diffp{\varPsi}{d} \, \dot d\\
                &=  g(d) \left[ \diffp{\psi}{\te F} \, : \dot{\te F}^\top 
                    + \sum\limits_{\varXi=1}^{2} \, \diffp{\, \, \psi}{\fviX} \, : {\fvidX}^\top \right]
                    + \diffp{g}{d}   \, \psi \, \dot d
                    \point
 \numberthis
 \label{eq:psiD}
 \end{align*}
 
 \subsubsection{Governing equations}
 The governing equations of the present model can be derived as necessary conditions for stationarity of the pseudo-potential $\Pilcd$ with respect to $\dot{\ve u}$, $\fvidX$ and $\dot d$,
 \begin{equation}
 \Pilcd \rightarrow \stat\limits_{{\dot{\ve u}, \fvidX, \dot d}}
 \qquad \Leftrightarrow \qquad
  \diffv{\Pilcd}{\, \dot{\ve u}} = \veZero \commam \diffv{\, \Pilcd}{\fvidX} = \te 0 \glmand \diffv{\Pilcd}{\, \dot{d}} = 0
  \comma
  \label{eq:varPrin} 
 \end{equation} 
which can be referred to as a rate type or incremental variational principle \cite{miehe2011}.%
\footnote{In the literature, the denomination \textit{incremental} variational principle is used for both time-continuous formulations of rate type as well as time-discrete approaches, see e.g. the overview of \cite{miehe2011}.}
Inserting the above definitions into \eqref{eq:varPrin}\textsubscript{2} and assuming sufficiently smooth field variables, which enables application of the divergence theorem, leads to the mechanical equilibrium condition
\begin{equation}
 \nabr \cdot \left(g(d) \, \diffp{\psi}{\te F} \right)^\top = \veZero 
 \quad \forall \, \ve X \in \omref
 \glmwith
 g(d) \, \diffp{\psi}{\te F} \cdot \ve N = \hat{\ve p} \quad \forall \, \ve X \in \domrefp
 \comma
 \label{eq:balLinMomPK1}
\end{equation} 
where $\ve N$ denotes the outward-pointing unit normal vector on $\partial \omref$, and in which the First Piola-Kirchhoff stress tensor $\te P$ and its virtually undamaged counterpart $\prescript{0}{}{\te P}$,
\begin{equation}
 \te P = g(d) \uP = g(d) \, \diffp{\psi}{\te F}
 \label{eq:stress}
\end{equation} 
can be identified.  The stress may be additively decomposed into the contributions arising from the individual over-stress branches,
\begin{equation}
  \uP = \sum\limits_{\varXi=1}^2 \, \uPoX 
  \qquad\glmwith\qquad 
  \uPoX = \diffp{\psistoX}{\,\, \te F}
  \point
\end{equation} 
Likewise, from \eqref{eq:varPrin}\textsubscript{3}, the evolution of viscous deformation is obtained as
\begin{equation}
 g(d) \, \diffp{\, \, \psi}{\fviX} + \diffp{\, \, \Phivd}{\fvidX} = \te 0
 \comma
 \label{eq:biot} 
\end{equation} 
which is a classical \textit{Biot type} evolution equation as considered in the framework of \textit{generalised standard materials}~\cite{biot1965,halphen1975}.
Making use of the above kinematic relations and after some manipulations, this relation can be rewritten in the form
\begin{equation}
  \utaoX = \viscX \, : \dvieX \comma
 \label{eq:visEvol_final} 
\end{equation}
which is similar to the original definition of finite viscoelasticity by Reese and Govindjee \cite{reese1998}, and resembles the classical evolution law of linear viscoelasticity,
with the \textit{virtually undamaged} Kirchhoff over-stress in branch $\varXi$ given by
\begin{equation}
 \utaoX = \uPoX \cdot \te F^\top \point
\end{equation} 
Finally, \eqref{eq:varPrin}\textsubscript{4} can be rewritten in a form similar to \eqref{eq:biot},
\begin{equation}
 \diffp{\varPsi}{d} + \diffv{(\Phifd+\Phifvd)}{\, \dot d} = 0 \comma
\end{equation} 
and the \textit{Ginzburg-Landau} type phase-field equation and the respective boundary conditions
\begin{equation}
   - \etaf \, \dot{d} = \diffp{g}{d} \, \psi
  + \gc \left( \frac{1}{2 \, \lc} d - 2\, \lc \, \nabr \cdot \nabr d \right)
  \quad \forall \, \ve X \in \omref
  \glmwith
  \nabr d \cdot \ve N  = 0 
  \quad \forall \, \ve X \in \partial \omref
   \label{eq:pfgl-unmod} 
\end{equation} 
are obtained.
Here, two non-variational extensions of the phase-field evolution equation \eqref{eq:pfgl-unmod}\textsubscript{1} are adopted.
\begin{figure}[tbp!]
		\centering 
		{\includegraphics[width=0.32\linewidth,trim=15mm 0mm 0mm 5mm]{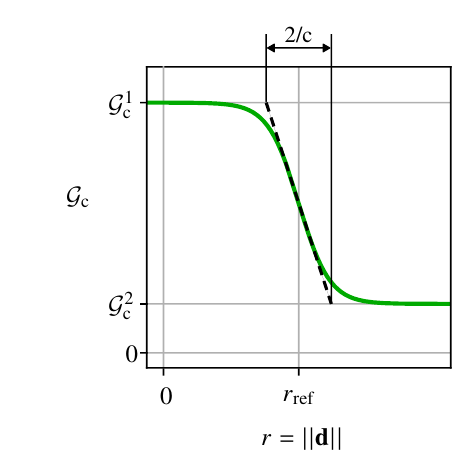}} % [trim=left bottom right top]
		\caption{Rate-dependent fracture resistance $\gc(\te d)$\textemdash Phenomenological ansatz}
		\label{fig:gcFun}
\end{figure}
Firstly, in order to enable the model to capture more complex material responses, a rate-dependent fracture resistance $\gc$ is introduced. Specifically, $\gc$ is allowed to depend on
\begin{equation}
 r (\te d) = \left\Arrowvert \te d \right\Arrowvert_\nv{F} = \sqrt{\te d \, \colon \te d} \comma
\end{equation}
which serves as a scalar measure of the rate of deformation.
In line with \cite{miehe2015b,dammass2023}, the sigmoid-shaped ansatz
\begin{equation}
 \gc(\te d) = \frac{\gco + \gct}{2} + \left( \frac{\gct - \gco}{2} \right)
 \, \tanh \left[ c \cdot (r(\te d) - \rref )\right]
 \label{eq:rabGc-def}
\end{equation}
is adopted as shown in Fig.~\ref{fig:gcFun}.
Secondly, in order to prevent \textit{crack healing} in case of unloading or an according change in rate of deformation, following the approach of Miehe et al.~\cite{miehe2010},
the history variable
\begin{equation}
 \hist = \max_{\tau \in [0,t]} \left\{ \frac{1}{\gc[\te d(\tau)]} \, \psi(\tau) \right\}
 \label{eq:histvar}
\end{equation} 
is introduced, which comprises the maximum of fracture driving force that has occurred.%
\footnote{By means of $\hist$, $\dot d \geq 0$ is exactly fulfilled in the absence of gradients, i.e. $\nabr d = \veZero$ \cite{miehe2010a}. For the heterogeneous case, to the best of the authors' knowledge, no analytical proof is available. Nevertheless, there is strong numerical evidence that $\dot d \geq 0$ is also fulfilled.}
Inserting these definitions, the phase-field equation \eqref{eq:pfgl-unmod}\textsubscript{1} is rewritten to
\begin{equation}
  - \frac{\etaf}{\gc} \, \dot{d} = \diffp{g}{d} \, \mathcal H 
  +  \left( \frac{1}{2 \, \lc} d - 2\, \lc \, \nabr \cdot \nabr d \right)
  \point
  \label{eq:pfgl-mod}
\end{equation}

\subsubsection{Thermodynamic consistency}
In order to ensure compatibility of the model with the second law of thermodynamics, following Coleman and Gurtin~\cite{coleman1967}, the Clausius-Duhem inequality
\begin{equation}
 \disd = \te P \, : \dot{\te F}^\top - \dot \varPsi \geq 0
 \label{eq:dissIneq}
\end{equation} 
shall hold for arbitrary $\dot{\te F}$, where $\dot \vartheta = 0$ and $\nabr \vartheta = \veZero$, i.e. isothermal conditions are assumed, here.
Inserting $\dot \varPsi$ according to~\eqref{eq:psiD} and $\te P$ from \eqref{eq:stress}, this constraint reduces to
\begin{equation}
\disd = \dvid + \dpfd \geq 0 \comma
\end{equation} 
with the viscous dissipation given by
\begin{equation}
 \dvid =  \sum\limits_{\varXi=1}^{2} - g(d) \, \, \diffp{\, \, \psi}{\fviX} \, : {\fvidX}^\top 
 = \sum\limits_{\varXi=1}^{2} \, \dvidX
 \label{eq:dvid-allg} 
\end{equation}
and the fracture dissipation
\begin{equation}
 \dpfd = \diffp{g}{d} \, \psi \, \dot d
 \point
\end{equation}
Since $\psi \geq 0$ and due to \eqref{eq:degfu-bed}\textsubscript{4}, $\dpfd \geq 0$ holds for $\dot d \geq 0$, which is ensured by means of the history variable approach as introduced in  \eqref{eq:histvar}--\eqref{eq:pfgl-mod}.
Moreover, inserting the \textit{Biot type} equation \eqref{eq:biot} into $\dvidX$ \eqref{eq:dvid-allg} and making use of the relation
\begin{equation}
 \diffp{\dvieX}{\fvidX} \, :  {\fvidX}^\top = \dvieX
 \comma
\end{equation}
the viscous dissipation within a single over-stress branch can be rewritten as
\begin{equation}
 \dvidX = \diffp{\, \, \Phivd}{\dvieX} \, : {\dvieX}^{\top} \comma
\end{equation} 
which is non-negative by definition of $\Phivd$ according to \eqref{eq:vis-dissP}.

\subsection{Finite element implementation aspects}
\begin{table}[b]
\caption{Governing equations for the present model considering the total Lagrangian approach: 
 Mechanical equilibrium~(a), phase-field equation~(b), viscous evolution~(c)}
\centering
\small
 \begin{tabular}{p{0.9\linewidth}}
 \hline
 \[ \nabr \cdot \left(\te F \cdot \te T \right)^\top = \veZero \tag{a} \] \\
{\begin{align*} 
  - \frac{\etaf}{\gc} \, \dot{d} =& \diffp{g}{d} \, \mathcal H 
  +  \left( \frac{1}{2 \, \lc} d - 2\, \lc \, \nabr \cdot \nabr d \right) \tag{b} 
\quad \text{with} \quad \hist = \max_{\tau \in [0,t]} \left\{ \frac{1}{\gc[\te d(\tau)]} \, \psi(\tau) \right\}
\end{align*}} \\
 \[ \utaoX = \viscX \, : \dvieX 
 \quad \text{with} \quad  
 \dvieX = \felX \cdot \fvidX \cdot {\te F}^{-1} 
 \quad \text{and} \quad
 \te F = \felX \cdot \fviX
 \tag{c}\] \\
  \hline
  \end{tabular}
%   }
\label{tab:gov-eqs} 
\end{table}

For the implementation of the model in a standard finite element code, the total Lagrangian approach is pursued. Accordingly, the mechanical equilibrium condition \eqref{eq:balLinMomPK1} is written in terms of the second Piola-Kirchhoff stress~\mbox{$\te T = \te F^{-1} \cdot \te P$},
\begin{equation}
 \nabr \cdot \left(\te F \cdot \te T \right)^\top = \veZero 
 \quad \forall \, \ve X \in \omref
 \quad \glmwith \quad 
 \te F \cdot \te T \cdot \ve N = \hat{\ve p} \quad \forall \, \ve X \in \domrefp
 \point
 \label{eq:balLinMomPK2}
\end{equation}
With this form of the mechanical equilibrium at hand, the governing differential equations of the present model are summarised in Tab.~\ref{tab:gov-eqs}.
For the derivation of the weak forms of the mechanical equilibrium \eqref{eq:balLinMomPK2} and the phase-field equation \eqref{eq:pfgl-mod}, the test function spaces
\begin{equation}
 \wus = \left\lbrace \wu \in \mathbb{H}^1(\omref) \,  \left| \, \wu = 0 \; 
  \forall \, \ve X \in \domrefu \right. \right \rbrace
  \quad \glmand \quad
  \wcs = \mathbb{H}^1(\omref)
\end{equation}
are defined, wherein $\mathbb{H}^1(\omref)$ is the Sobolev space of square integrable functions possessing square integrable derivatives in $\omref$, and $\domrefu = \domref \setminus \domrefp$ denote the parts of the boundary where the respective components of the displacement vector $\ve u$ are prescribed.
Following the method of weighted residuals with $\wu \in \wus$ and $\wc \in \wcs$, and making use of integration by parts and the divergence theorem yields
\begin{equation}
  \inte{\omref}{\left(\te F \cdot \te T \right)^\top \, \colon \left(\nabr \wu\right)^\top}{V} 
  - \inte{\domrefp}{\hat{\ve p} \, \wu}{A} = 0 \comma
  \label{eq:impb-s}
\end{equation} 
and
\begin{equation}
  \inte{\omref}{ \left( \diffp{g}{d} \, \mathcal H  + \frac{1}{2 \, \lc} \, d + \frac{\etaf}{\gc} 
 \, \dot d \right) \, \wc 
  + 2 \, \lc \, \nabr d \cdot \nabr \wc }{V} 
 = 0
 \point
 \label{eq:pfgl-s}
\end{equation} 
For spatial discretization, Galerkin's method is applied and time discrete forms are obtained by approximating the respective rates using an Euler backward scheme.

In order to integrate the viscous evolution equation \eqref{eq:visEvol_final}, the rate of the elastic stretch tensors 
\begin{equation}
 \beldX
 =
 \underbrace{\te l \cdot \belX + \belX \cdot \te l^\top}_\nv{predictor}  \, 
%  + \, \underbrace{\te F \cdot \left[ \left(\cviXold\right)^{-1} \right]^{\boldsymbol{\cdot}} \cdot \te F^\top}_\nv{corrector}
+ \, \underbrace{\lie\left[ \belX \right]}_\nv{corrector}
 \label{eq:opSplit}
\end{equation}
is approximated by means of a \textit{predictor-corrector} type operator split algorithm.
Based on the inelastic deformation of the previous time step $t_{n-1}$ and the current deformation gradient $\te F_n$, a \textit{trial} value of elastic stretch
\begin{equation}
 \beltrnX = \te F_n \cdot \left(\cviXold\right)^{-1} \cdot \te F^\top_n
\end{equation} 
is defined in the \textit{predictor} step. Subsequently, a \textit{corrected} value of $\belnX$ that accounts for the change in inelastic deformation is determined, for which $\te l = \te 0$ is set and the evolution equation~\eqref{eq:visEvol_final} together with the kinematic relation~\eqref{eq:kinRel_lie_dvi} are inserted into~\eqref{eq:opSplit}.
The resulting differential equation
\begin{equation}
 \beldX = -2 \, \left( \viscX^{-1} \, \colon \utaoX \right) \cdot \belX 
\end{equation} 
is integrated in approximate manner by means of an exponential mapping scheme, which leads to a nonlinear algebraic system of equations that is iteratively solved for $\belX$ by means of a \textit{local} Newton scheme at quadrature point level in each iteration step of the \textit{global} finite element solution procedure.
More details on the algorithmic treatment of viscous evolution are provided in the previous contribution \cite{dammass2023}, where special attention is also paid to the plane stress case, and the fundamental work \cite{reese1998}.

The coupled problem is solved by means of a \textit{staggered} approach, i.e. the mechanical equilibrium and the phase-field equation are solved in sequential manner.
Multiple iterations of this decoupled solution scheme are performed until a convergence criterion is fulfilled, which is based on the update of the nodal degrees of freedom with respect to the previous \textit{staggered} loop iteration.
Furthermore, adaptive control of the time step size is employed based on a heuristic scheme. For this, depending on the number of \textit{global} Newton iterations as well as on the number of staggered loops, the time step is either lowered or increased.
In addition, if the convergence criteria are not met after a certain number of Newton iterations or \textit{staggered} loops, respectively, a \textit{cut back} to the last converged increment is performed and the time step size is lowered, accordingly.

\section{Material and specimen preparation}
\label{sec:mat}

Generally, caramel is a class of confectionery that is made from various sugars, water, milk solids, fats, and emulsifiers. After mixing, homogenisation and heating to temperatures between $\SI{118}{\degree C}$ and $\SI{130}{\degree C}$, food products are obtained, which feature a content of moisture typically ranging from \mbox{6 g/100 g} to \mbox{12 g/100 g} \cite{ahmed2006,hartel2018}.

\begin{figure}[tbp!]
		\centering 
		{\includegraphics[width=0.75\linewidth,trim=0mm 5mm 0mm 10mm]{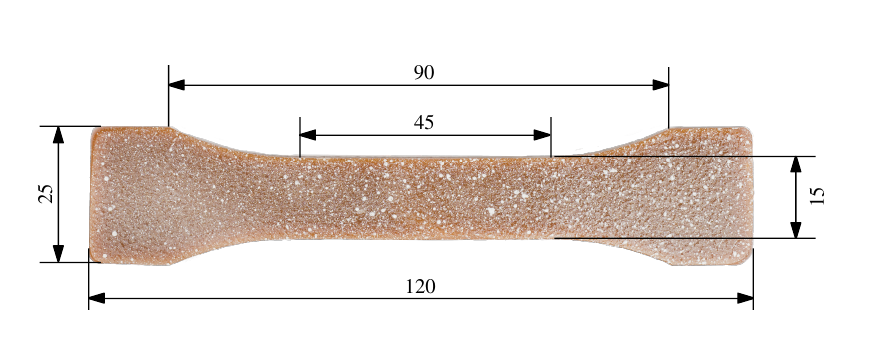}} % [trim=left bottom right top]
		\caption{\textit{Dog bone}-shaped specimens used for the investigation of the deformation behaviour. The dimensions are given in millimetres, and the thickness is 5~mm. In order to enable Digital image correlation, the white stochastic pattern of acrylic paint is sprayed on the specimens.}
		\label{fig:dogbone}
\end{figure}

Within the scope of this contribution, a toffee-like caramel is considered that is produced under lab conditions with the respective procedure comprehensively described in the previous work~\cite{schab2021}.
The corresponding premix consists of granulated sucrose, glucose syrup, water, skim milk powder, palm oil, palm stearin and soy lecithin. 
After mixing these ingredients, the heating process is performed with ongoing stirring and temperature monitoring  until a temperature of $\SI{120.5}{\degree C}$ is reached. At this point, the cooking procedure is terminated and the mass is transferred into dog-bone shaped molds, with the respective dimensions shown in Fig.~\ref{fig:dogbone}.

The resulting toffee material has a content of moisture of \mbox{(7.0  g ± 0.2 g)/100 g}, which is verified for each lot of specimens by means of gravimetric analysis. Apart from that, it consists of \mbox{39.8 g/100 g} sucrose, \mbox{36.5 g/100 g} non-sucrose saccharides, \mbox{10.6 g/100 g} fat, \mbox{3.9 g/100 g} protein, \mbox{1.2 g/100 g} lecithin, and \mbox{1.0 g/100 g} mineral salts, with the mean mass fractions of these ingredients calculated from the gravimetrically-determined moisture content and the respective mass portions in the premix.
While this confectionery is macroscopically homogeneous, on the micro-scale, it can be considered as a multiphase system that consists of a continuous phase and a dispersed phase of fine fat droplets, cf. \cite{heathcock1985,sengar2014}.
The continuous phase mainly consists of amorphous sugar, in which a protein network is formed during heating. Moreover, it is in this phase where there is a significant content of moisture, and some fluid-like aspects in the mechanical behaviour of the material can be attributed to this, cf. \cite{miller2015}.

\section{Mechanical behaviour of toffee at room temperature}
\label{sec:mech-25}
For the experimental investigation of the complex rate- and temperature-dependent mechanical behaviour, a gradual approach is pursued. In this section, the case of the temperature being constant and equal to the room value $\vartheta=\SI{25}{\degree C}$ is considered. Furthermore, the deformation behaviour of the bulk is studied first, for which unnotched specimens are examined. Subsequently, fracture is investigated by way of pre-notched specimens.

\subsection{Deformation behaviour at room temperature}
\label{sec:defo-25}

\subsubsection{Experimental results}
For the experimental analysis of the deformation behaviour, \textit{dog bone} specimens are considered under tension as shown in Fig.~\ref{fig:dogbone}. A stochastic pattern of acrylic paint is sprayed on the specimen surface, which enables to identify the deformation field by means of Digital image correlation (DIC).
Plane stress conditions are assumed.
Displacement control is applied. In order to characterise the rate- or time-dependency, respectively, relaxation tests are carried out and monotonic experiments are performed at different rates.
For each specific testing condition, at least ten repetitions of  the respective experiment are performed.

\begin{figure}[tbp!]
		\centering 
		\subfloat[]{\includegraphics[width=0.45\linewidth,trim=0mm 0mm 0mm 0mm]{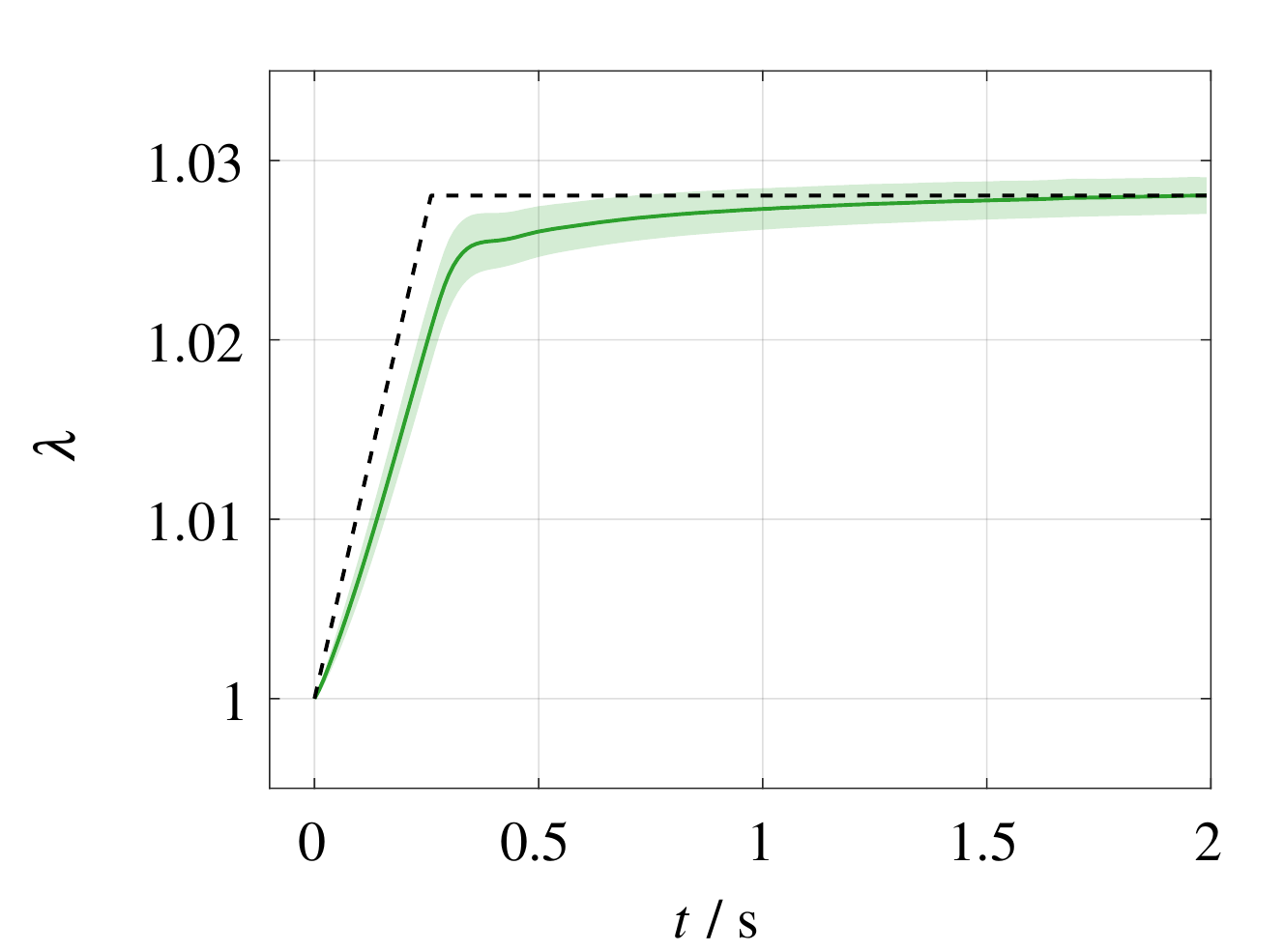}\label{fig:relax_25C_lam}} % [trim=left bottom right top]
		%Prescribed stretch\textemdash{}Actual experimental data vs. idealised curve prescribed to the testing machine
        \hspace{1.0cm}
        \subfloat[]{\includegraphics[width=0.45\linewidth,trim=0mm 0mm 0mm 0mm]{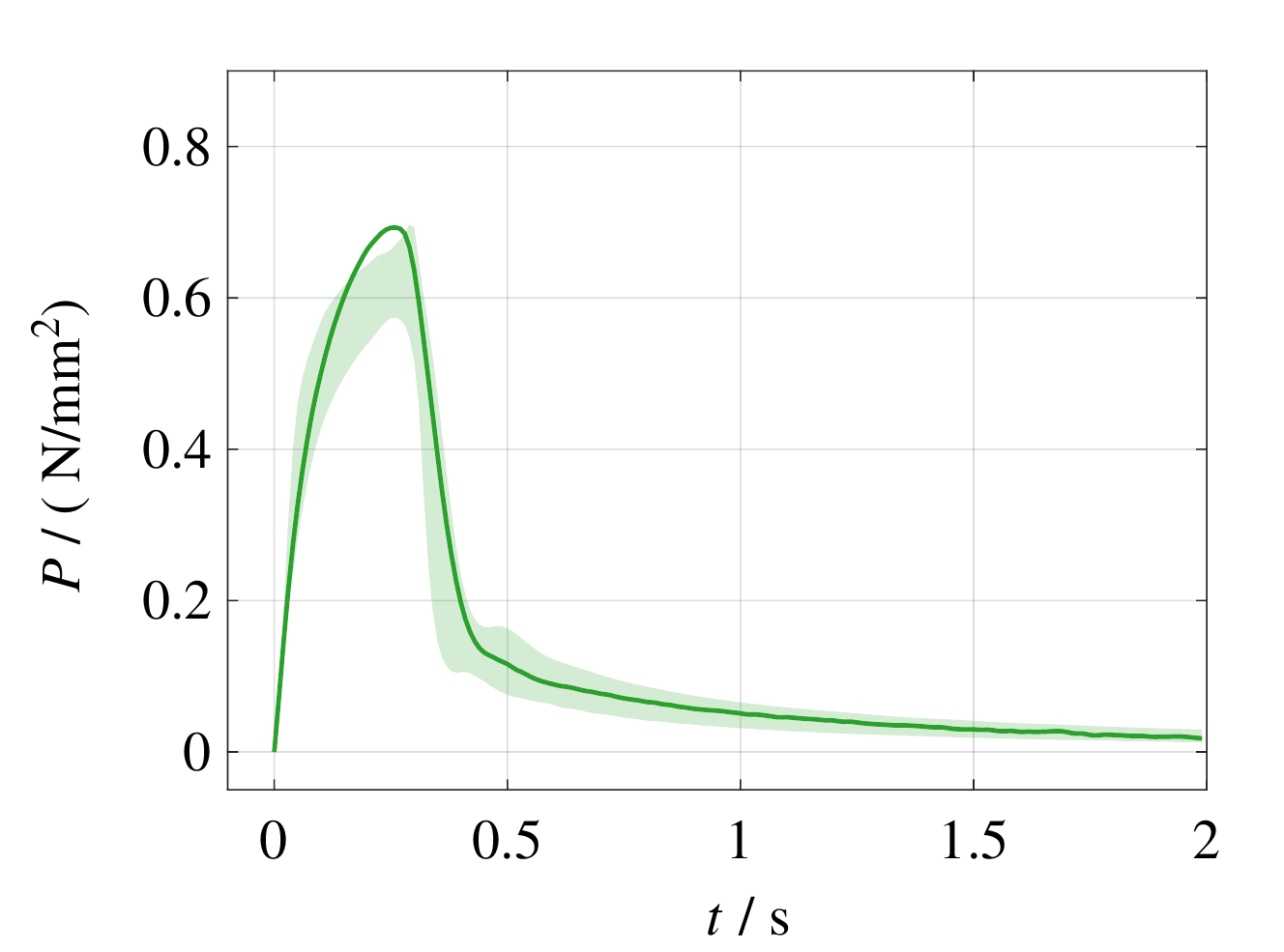} \label{fig:relax_25C_P}} \\% [trim=left bottom right top]
        %[Resulting stress]
        \includegraphics[scale=0.71,trim=0mm 5mm 15mm -2.5mm]{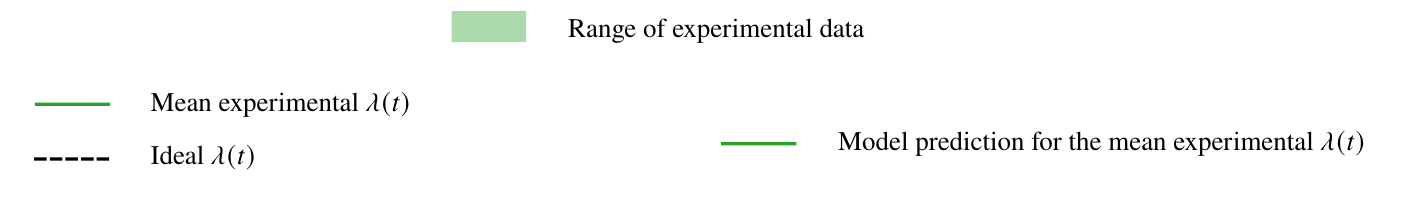}
		\caption{Stress relaxation within unnotched specimens at room temperature~($\vartheta = \SI{25}{\degree C}$). Actual experimental $\lambda$-$t$ data vs. ideal $\lambda$-$t$ curve (a), and the  experimental $P$-$t$ data compared to the corresponding prediction of the parameterised model for the mean experimental $\lambda(t)$~(b)}
		
		\label{fig:relax_25C}
\end{figure}
For the relaxation tests, the stretch-time data and the corresponding stress response are presented in Fig.~\ref{fig:relax_25C}.
As shown in Fig.~\ref{fig:relax_25C_lam}, a displacement ramp is prescribed until a certain value is reached, which is then hold constant. In order to avoid any relevant damage within these experiments, this value is chosen such that the stretch $\lambda$ within the zone of constant cross section in the centre of the specimen is rather small.
For both $\lambda$ and the corresponding nominal stress $P$, in Fig.~\ref{fig:relax_25C}, the scatter in the experimental data is accounted for by means of the 68.3~\%~confidence interval, which is estimated from the repetitive experiments in assuming Gaussian contributions for each incremental value of the respective quantities.%
\footnote{For time synchronisation of the stretch and stress data that stem from different repetitions of the experiments, piecewise spline interpolation is used for the incremental data, cf.~\cite[p. 44f.]{kalina2020}.}
In what follows, this interval is referred to as the range of experimental data.
As some scatter in experimental results is inevitable when dealing with food-like natural materials, deviation can be denominated comparably small for this specific experiment.

As it is shown in Fig.~\ref{fig:relax_25C_P}, when $\lambda$ remains at almost constant value, the stress $P$ rapidly decreases and approaches zero. Accordingly, viscous effects, which result in the time-dependency of the material response, are highly significant and the corresponding relaxation times seem to be small.
Furthermore, there obviously is no equilibrium stress, i.e. $P(t\rightarrow \infty) = 0$.
As a consequence, toffee can be classified as a viscous material or liquid, where the stress essentially is a function of stretch rate, cf.~\cite{boisly2014}.
Alternatively, it may also be suitable to classify the material either as a viscoelastic solid with vanishing equilibrium stiffness or as viscoplastic without a distinct yield stress, cf.~\cite{haupt1993}.

\begin{figure}[tbp!]
		\centering 
		{\includegraphics[width=0.475\linewidth,trim=0mm 0mm 0mm 0mm]{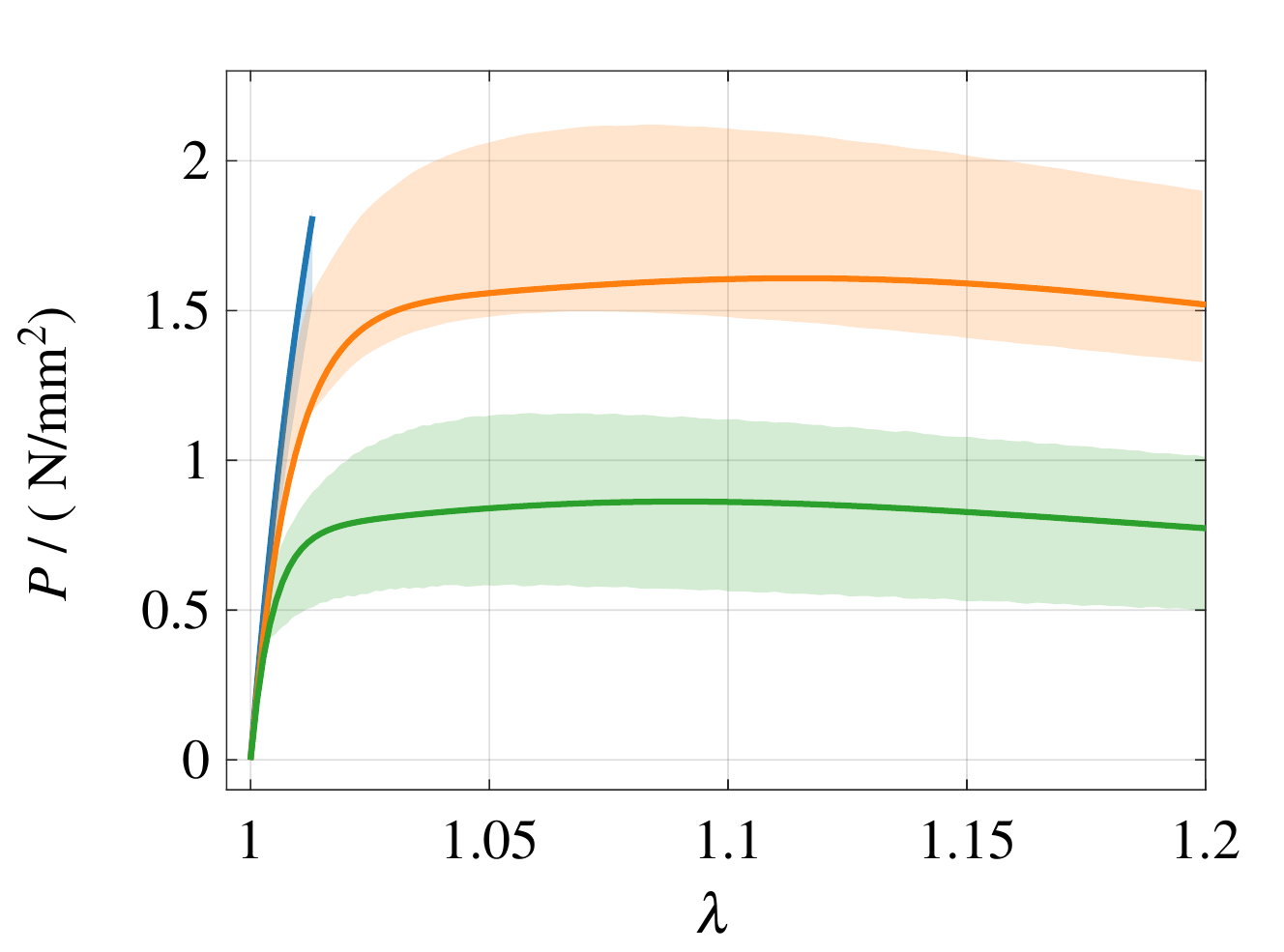}} % [trim=left bottom right top]
		\hspace{2mm}
        \includegraphics[scale=0.95,trim=0mm -9mm 0mm 0mm]{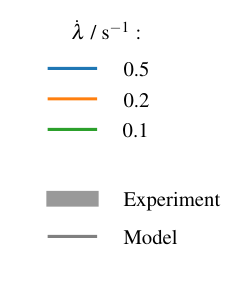}
		\caption{Monotonic tensile tests unnotched specimens at room temperature~($\vartheta = \SI{25}{\degree C}$). Experimental stress-stretch data vs. prediction of the parameterised model for different values of (nominal) stretch rate $\dot{\lambda}$}		
		\label{fig:defo_mon_25C}
\end{figure}
The $P$-$\lambda$ data from the monotonic experiments, Fig.~\ref{fig:defo_mon_25C}, also reveal a significant rate-dependency of the deformation behaviour.
In particular, for a given stretch $\lambda$, the stress $P$ increases with the prescribed rate $\dot{\lambda}$.
Furthermore, for low and intermediate rates, $P$ does no longer increase with $\lambda$, when a certain value of $\lambda$ is reached, yet approximately remains at a constant plateau value that seems to depend on the stretch rate $\dot \lambda$, only. 
For large stretches $\lambda \gtrsim 1.1$, even a slight decrease of stress with increasing stretch can be observed, which is true for both the nominal stress $P$ as shown in Fig.~\ref{fig:defo_mon_25C} as well as the Cauchy stress $\sigma$ (not depicted).%
\footnote{Since the stress response obviously is highly dissipative and decisively depends on rate of deformation, one reason for the stress decrease at higher stretches may be the decrease of the norm of the rate of deformation tensor $||\te d||$ with $\lambda$ given a constant $\dot \lambda$.
In addition, for larger stretches $\lambda \geq 1.2$, damage effects in the bulk material might also play a part, which go beyond the scope of the present contribution, though.}
Different from the case of small stretches, which has been considered for the relaxation tests, the scatter in the experimental data becomes very pronounced for larger deformations. Primarily, this effect can be attributed to slight variations in the exact chemical composition between the individual specimens and some microscopic heterogeneity, which are inevitable when dealing with a food-like natural material.
For high $\dot \lambda$, highly brittle failure of the specimens has been observed, which also is assumed to be triggered by these microscopic heterogeneities or defects. Accordingly, $P$-$\lambda$ data only is available for a small range of stretches in these cases.

\subsubsection{Parameter identification}

In order to identify the viscous parameters of the bulk and the belonging relaxation times, the deviation between mean experimental stress data from the monotonic and the relaxation tests and the according model prediction is minimised.
To this end, homogeneous uniaxial stress states are assumed and damage is not taken into account. 
Furthermore, since no relevant time-dependency of the transversal deformation has been observed in the relaxation tests, the Poisson's ratios are assumed to be identical for all over-stress branches and set to $\nuovnX = 0.47$, which corresponds to the mean value computed from the experimentally determined axial and transversal stretches.
Then, in order to identify the Ogden parameters $\muovX[p]{\varXi}$ and $\alovX[p]{\varXi}$ 
as well as the relaxation times $\tauX$, the \textit{GlobalSearch} framework of Matlab R2020b together with the \textit{fmincon} algorithm for constrained optimisation problems is employed.
In doing so, it is demanded that the constants satisfy the requirement
\begin{equation}
 \alovX[p]{\varXi} \, \muovX[p]{\varXi} \geq 0 
\end{equation}
for any $p$. 
Furthermore, for the relaxation times of each over-stress branch, an interval of admissible values is defined such that the individual $\tauX$ do approximately differ by one order of magnitude, where the lower bound for the smallest $\tauX$ is defined based on the temporal resolution of the experimental data and the upper bound for the largest $\tauX$ determined by the hold time of the relaxation experiments.
To a certain extent, this approach is inspired by the \textit{window algorithm} developed for the identification of relaxation spectra in the framework of linear viscoelasticity, see \cite{emri1995} for an overview.
A comparative study for generalised Maxwell elements of varied complexity has revealed that two non-equilibrium branches are sufficient for adequately capturing the experimental responses, whereas the accuracy of the model can not be remarkably  increased further by means of additional branches.
Likewise, for the number of Ogden terms, $\NovX=2$ has revealed adequate, which has been set to the identical value for the two over-stress branches for the sake of simplicity.
The identified parameters of the deformation model are listed in Tab.~\ref{tab:matpar-bulk_25}. From comparison of the model with the experimental stress data as shown in Figs.~\ref{fig:relax_25C_P} and \ref{fig:defo_mon_25C}, good agreement can be stated.
\begin{table}[tb]
    \caption{Parameters of the model of bulk deformation at $\vartheta = \SI{25}{\degree C}$}
    \centering
    \small
%     \resizebox{\linewidth}{!}{
    \begin{tabular}{*7c}
    \toprule
    $\varXi$ &
    $\nuovnX$ & $\muovX[1]{\varXi} / (\nv{N/mm^2})$ & $\alovX[1]{\varXi}$ & $\muovX[2]{\varXi} / (\nv{N/mm^2})$ & $\alovX[2]{\varXi}$ & $\tauX / \nv{s}$ \\
    \midrule
    1 & \multirow{2}{*}{0.47} & 99752.1 & 0.00129 & 0.0005 & 19.73 & 0.039 \\[3pt]
    2 & & 0.1176 & 20 & 0 & -- & 1 \\
        \bottomrule
    \end{tabular}
%     }
    \label{tab:matpar-bulk_25}
\end{table}

\subsection{Rate-dependent fracture behaviour at room temperature}
\label{sec:frac-25}

\subsubsection{Fracture parameter identification\textemdash Single-notched specimens}
In order to parameterise the rate-dependent fracture resistance function $\gc(\te d)$, specimens of similar geometry are considered as for the investigation of the deformation behaviour, see Fig.~\ref{fig:dogbone}. In order to ease the investigation of crack propagation, notches of 7~mm length are generated in the centre of the specimens using razor blades.
With these single-notched specimens, monotonic tensile tests (\textit{SENT}) are performed at varied displacement rates $\urbd \in [500, 2500] \, \si{mm/min}$.

\begin{figure}[tbp!]
		\centering
        \subfloat[Brittle fracture (\textit{here:} $\urbd = \SI{2500}{mm/min}$, $\vartheta=\SI{25}{\degree C}$)]
        {\includegraphics[scale=1.0,trim=0mm 0mm 0mm 0mm]{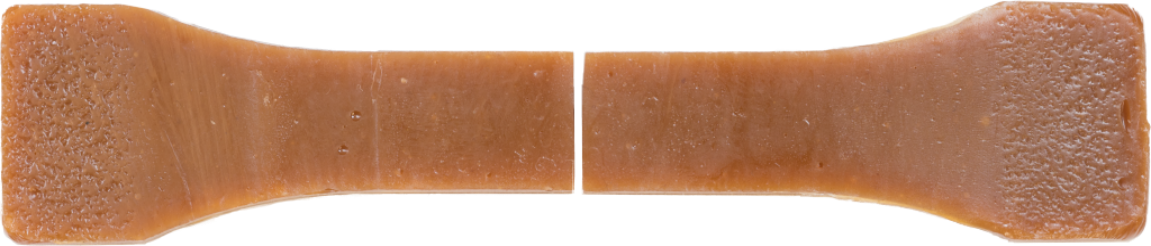}\label{fig:sent_exp_brit}} % [trim=left bottom right top]
        
        \vspace{3mm}
		
		\subfloat[Ductile fracture (\textit{here:} $\urbd = \SI{800}{mm/min}$, $\vartheta=\SI{25}{\degree C}$; crack path highlighted by blue line)]
		{\includegraphics[scale=1.0,trim=0mm 0mm 0mm 0mm]{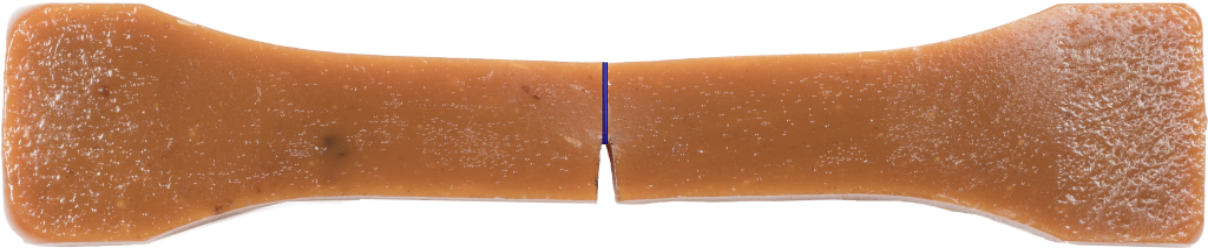}\label{fig:sent_exp_ductFrac}} % [trim=left bottom right top]
		
		\vspace{3mm}
		
        \subfloat[Ductile behaviour without crack propagation (\textit{here:} $\urbd = \SI{500}{mm/min}$, $\vartheta=\SI{25}{\degree C}$)]{\includegraphics[scale=1.0,trim=0mm 0mm 0mm 0mm]{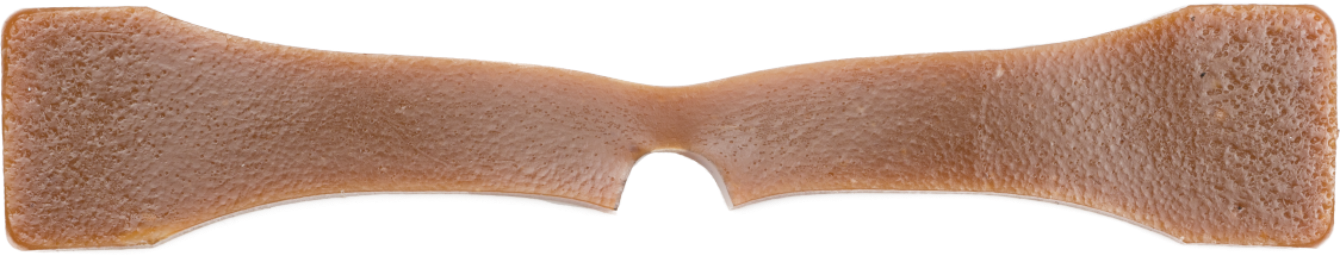}\label{fig:sent_exp_duct}} % [trim=left bottom right top]
		
		\caption{Single-notched specimens under tension with varied displacement rates $\urbd$~(\textit{here:} $\vartheta = \SI{25}{\degree C}$){}\textemdash Experimental results}		
		\label{fig:sent_exp}
\end{figure}

\begin{figure}[tbp!]
		\centering 
		{\includegraphics[width=0.5\linewidth,trim=0mm 0mm 0mm 0mm]{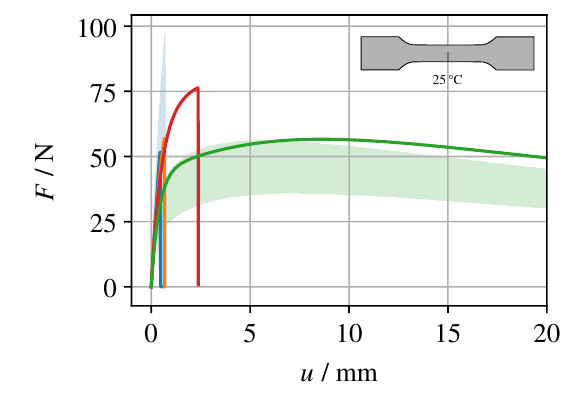}} % [trim=left bottom right top]
		\hspace{2mm}
        \includegraphics[scale=0.9,trim=0mm -15mm 0mm 5mm]{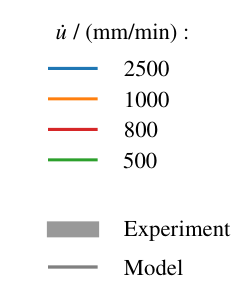}
		\caption{Monotonic tensile tests with single-notched specimens at room temperature~($\vartheta = \SI{25}{\degree C}$).
		Experimental force-displacement data compared to the model.
		For $\dot u = \SI{800}{mm/min}$, i.e. for an intermediate displacement rate in the range where the transition between  ductile and highly behaviour, experimentally observed phenomena are ambiguous to a certain extent, which is why the corresponding simulated curve is shown, only.}		
		\label{fig:sent_fu_25C}
\end{figure}

Depending on the prescribed value of $\urbd$, the experimental results differ drastically. Essentially, three phenomena can be distinguished, which are shown in Fig.~\ref{fig:sent_exp}. The corresponding $F$-$u$ curves are given in Fig.~\ref{fig:sent_fu_25C}.
Firstly, for \textit{high} $\urbd \gtrsim \SI{1000}{mm/min}$, \textit{highly brittle} responses are observed with a crack initiating at the tip of the pre-existing notch and instantaneously propagating towards the opposite edge of the specimen at $u \lesssim \SI{1}{mm}$, Fig.~\ref{fig:sent_exp_brit}.
Secondly, for \textit{low} $\urbd \lesssim \SI{500}{mm/min}$, it does not come to crack propagation at all. Instead, the specimens can be \textit{excessively deformed} in inelastic manner, Fig.~\ref{fig:sent_exp_duct}.
Thirdly, for intermediate rates, ductile fracture can be observed, which is preceded by a significant amount of inelastic deformation, Fig.~\ref{fig:sent_exp_ductFrac}.
However, for these rates within the \textit{transition range} between highly brittle fracture and ductile deformation without crack propagation, experimental results have revealed ambiguous. More precisely, for the \textit{SENT} repeated at $\urbd = \SI{800}{mm/min}$ and $\urbd = \SI{750}{mm/min}$, all the three aforementioned phenomena have been observed.
Most likely, this can be attributed to some uncertainty in the exact composition of the individual specimens such as slight deviations in moisture content. Moreover, regarding the micro-scale, there are small air bubbles and dispersed fat droplets.
When performing experiments with natural materials such as food, these stochastically distributed micro-heterogeneities are inevitable. They can be assumed to introduce some scatter in the experimental responses.

\begin{figure}[tbp!]
		\centering
        \subfloat[Brittle fracture (\textit{here:} $\urbd = \SI{2500}{mm/min}$, $\vartheta=\SI{25}{\degree C}$)]
        {\includegraphics[scale=0.175,trim=0mm 0mm 0mm 0mm]{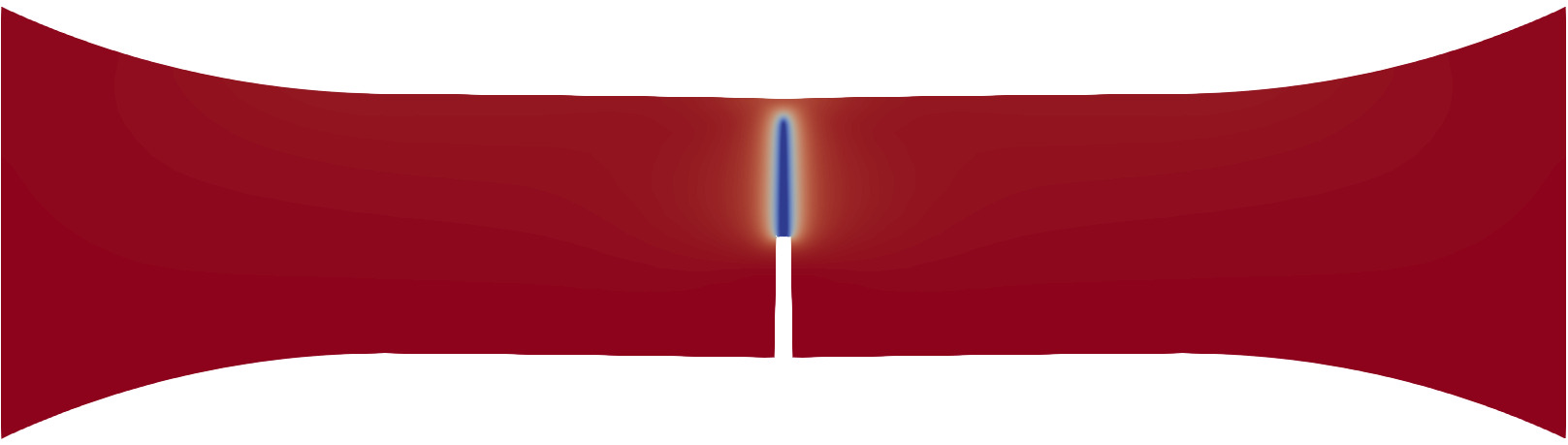}\label{fig:sent_num_brit}} % [trim=left bottom right top]
        
        \vspace{3mm}
		
		\subfloat[Ductile fracture (\textit{here:} $\urbd = \SI{800}{mm/min}$, $\vartheta=\SI{25}{\degree C}$)]
		{\includegraphics[scale=0.175,trim=0mm 0mm 0mm 0mm]{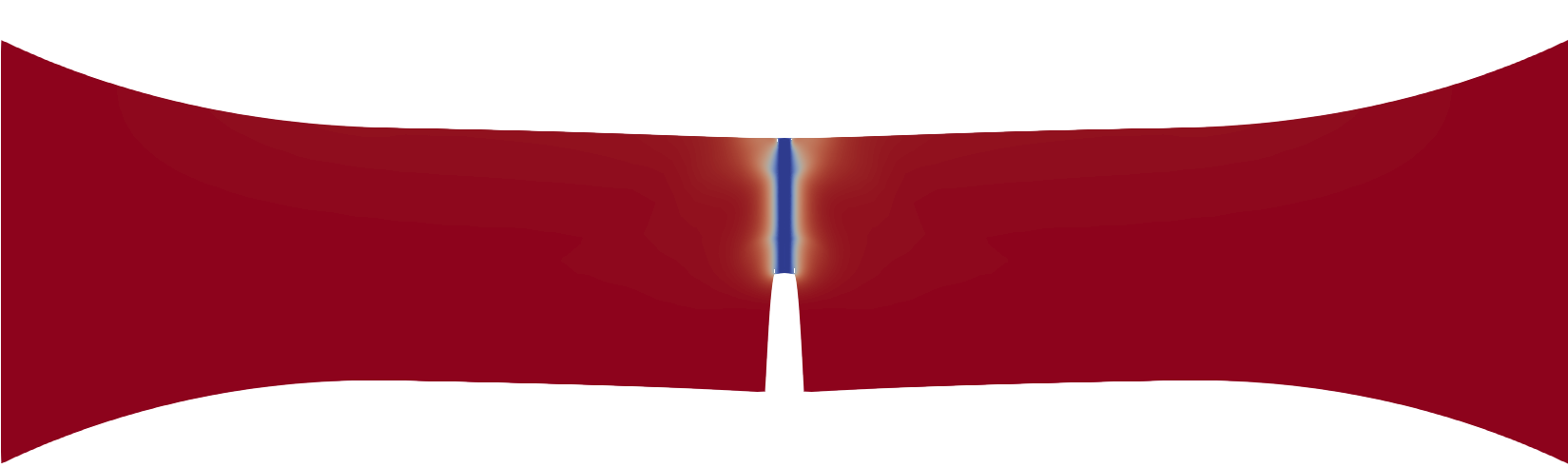}\label{fig:sent_num_ductFrac}} % [trim=left bottom right top]
		
		\vspace{3mm}
		
        \subfloat[Ductile behaviour without crack propagation (\textit{here:} $\urbd = \SI{500}{mm/min}$, $\vartheta=\SI{25}{\degree C}$)]{\includegraphics[scale=0.175,trim=0mm 0mm 0mm 0mm]{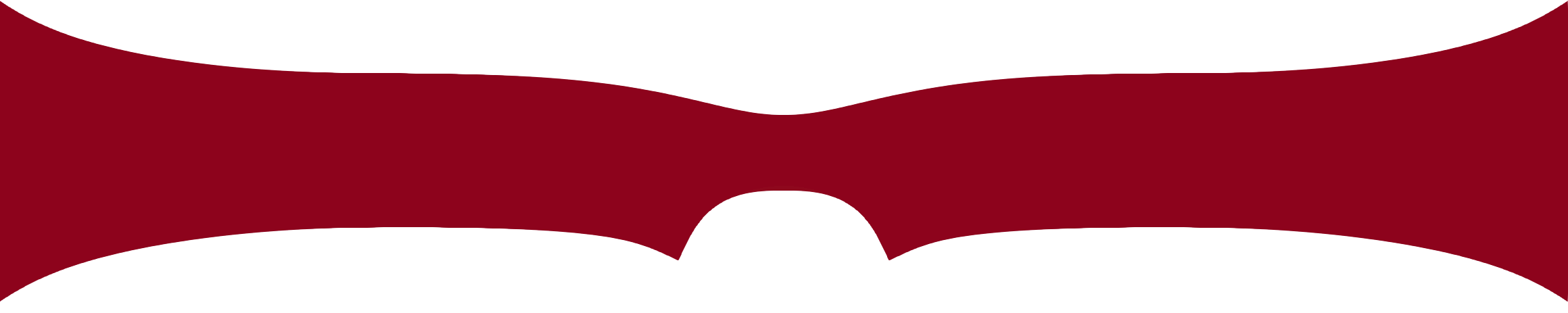}\label{fig:sent_num_duct}} % [trim=left bottom right top]
        
        \vspace{3mm}
        \includegraphics[scale=0.38,trim=0mm 0mm 0mm 0mm]{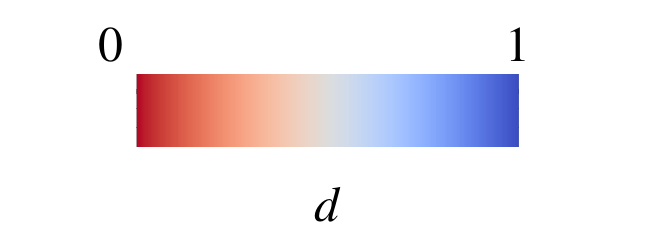}% [trim=left bottom right top]
		
		\caption{Single-notched specimens under tension with varied displacement rates $\urbd$~(\textit{here:} $\vartheta = \SI{25}{\degree C}$){}\textemdash Numerical results}		
		\label{fig:sent_num}
\end{figure}
For the parameterisation of the rate-dependent fracture resistance $\gc(\te d)$, the mean experimental $F$-$u$ curves from the \textit{SENT} are compared to model predictions with varied parameters of the phase-field model, while the model of deformation behaviour is a priori fixed.
In doing so, the length scale parameter $\lc$ is predefined to $\lc=\SI{0.25}{\milli\meter}$, which is considerably smaller than the minimal dimension of the specimen while allowing for an acceptable number of elements required for resolving the diffuse crack topology.
Moreover, the numerically motivated kinetic fracture parameter and the residual stiffness are chosen to $\etaf=\SI{5e-6}{Ns\per\milli\meter^2}$ and $k=10^{-10}$, respectively.
For all the simulations presented in this paper, the pre-existing cuts within the specimens are modelled geometrically by means of notches.
Since the specimens are of low thickness, plane stress conditions are assumed and two-dimensional simulations are performed, here.
Due to symmetry, only one half of the specimen domain is considered. The mesh consists of quadratic triangular elements and is refined along the expected crack path. By means of a numerical comparison of different mesh sizes, $h$-convergence is verified.
The constants $\gco$, $\gct$, $c$ and $\rref$ are then identified such that the mean experimental critical displacement $u_\nv{crit}$ is optimally recovered in case of brittle fracture for $\dot u \geq \SI{1000}{mm/min}$, while it does not come to crack propagation when $\dot u \leq \SI{500}{mm/min}$.
Moreover, for the \textit{transition range} between brittle fracture and crack propagation-free deformation, an assumption is made. While the experimental results are somewhat ambiguous, only ductile fracture is considered at the mean $u_\nv{crit}$ observed in the experiments for the parameterisation.%
\footnote{In order to reproduce the ambiguousness of the experimental results for rates in the \textit{transition range} with the model, it would be necessary to explicitly account for the stochasticity in the material properties. To this end, a stochastic fracture phase-field modelling could be carried out, cf.~\cite{gerasimov2020,nagaraja2023a}, which, however, goes beyond the scope of the present publication.}
The final parameters are listed in Tab.~\ref{tab:matpar-pf_25}, and as it becomes clear from Fig.~\ref{fig:sent_num}, the parameterised model is capable of capturing all the three experimental results in terms of crack pattern or deformed geometry, respectively.
Furthermore, as shown in Fig.~\ref{fig:sent_fu_25C}, good agreement between numerical and experimental $F$-$u$ data is obtained, which also serves as a validation of the model bulk deformation with its parameters identified as discussed in Sect.~\ref{sec:defo-25}.
\begin{table}[tb]
    \caption{Parameters of the fracture phase-field for $\vartheta = \SI{25}{\degree C}$}
    \centering
    \small
\begin{tabular}{*7c}
    \toprule
    $\gco / (\nv{N/mm})$ & $\gct / (\nv{N/mm})$ & $c / \nv{s}$ & $\rref / \nv{s}^{-1}$   & $ \etaf / (\nv{N s}/\nv{mm}^2)$  & $k$ & $\lc / \nv{mm}$ \\
    \midrule
    16 & 0.15 & 20 & 2.15 & $5 \cdot 10^ {-6}$ & $10^{-10}$ & 0.25 \\
    \bottomrule
        
\end{tabular}
    \label{tab:matpar-pf_25}
\end{table}

\subsubsection{Model validation\textemdash Double-notched specimens}

\begin{figure}[tbp!]
		\centering
        \subfloat[Brittle fracture (\textit{here:} $\urbd = \SI{2500}{mm/min}$, $\vartheta=\SI{25}{\degree C}$; crack path highlighted by blue line)]
        {\includegraphics[scale=1.0,trim=0mm 0mm 0mm 0mm]{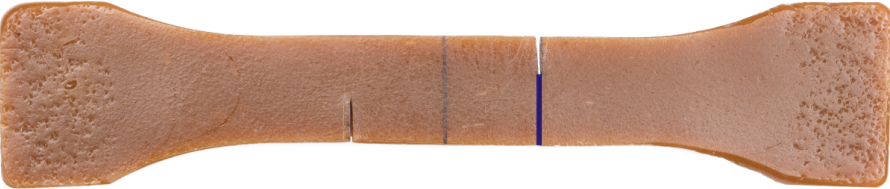}\label{fig:dent_exp_brit}} % [trim=left bottom right top]
        
		\vspace{3mm}
		
        \subfloat[Ductile behaviour without crack propagation (\textit{here:} $\urbd = \SI{500}{mm/min}$, $\vartheta=\SI{25}{\degree C}$)]{\includegraphics[scale=1.0,trim=0mm 0mm 0mm 0mm]{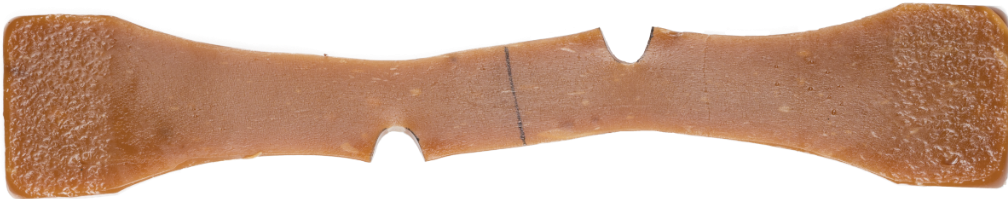}\label{fig:dent_exp_duct}} % [trim=left bottom right top]
		
		\caption{Double-notched specimens under tension with varied displacement rates $\urbd$~(\textit{here:} $\vartheta = \SI{25}{\degree C}$){}\textemdash Experimental results}		
		\label{fig:dent_exp}
\end{figure}

\begin{figure}[tbp!]
		\centering 
		{\includegraphics[width=0.5\linewidth,trim=0mm 0mm 0mm 0mm]{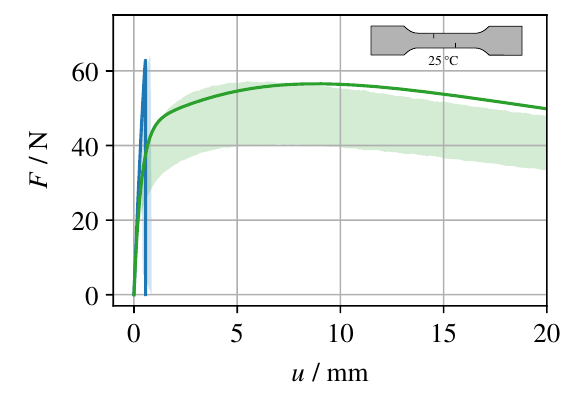}} % [trim=left bottom right top]
		\hspace{2mm}
        \includegraphics[scale=0.9,trim=0mm -25mm 0mm 5mm]{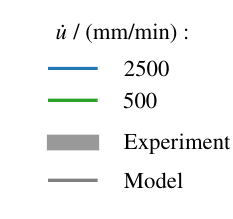}
		\caption{Monotonic tensile tests with double-notched specimens at room temperature~($\vartheta = \SI{25}{\degree C}$).}		
		\label{fig:dent_fu_25C}
\end{figure}

With the aim of validating the entire model, asymmetrically double-notched specimens are investigated under tension~\textit{(DENT)} at varied $\urbd \in [500, 2500] \, \si{mm/min}$.
To this end, the same dog bone-shaped specimens are considered as in the previous sections, Fig.~\ref{fig:dogbone}, and, at a distance of $\SI{12.5}{mm}$ from the centre, two straight vertical notches of $\SI{5}{mm}$ length are introduced using a razor blade, with one starting at the bottom edge and the other one from the top edge.
As shown in Figs.~\ref{fig:dent_exp} and \ref{fig:dent_fu_25C}, the experimental responses are similar between \textit{DENT} and \textit{SENT} regarding the rate-dependency. Whereas specimens deform without crack propagation for low rates of the prescribed displacement $\dot u$, Fig.~\ref{fig:dent_exp_duct}, highly brittle fracture is observed for high $\dot u$, Fig.~\ref{fig:dent_exp_brit}. Again, ambiguous results are obtained for intermediate rates, which is attributed to microscopic heterogeneity and small deviations in the chemical composition between the individual specimens.
When it comes to fracture, starting at one of the notch tips, a crack instantaneously propagates through the specimen almost straightly towards the opposite vertical edge. 
When repeating the \textit{DENT}, cracks start from either the right or the left edge, while in each case only one crack is forming.
This crack pattern is similar to experimental observations for viscoelastic materials that are considered for technical applications, e.g. Hydroxyl-terminated polybutadiene (HTPB)-based solid propellants \cite{han2012}. However, when performing the \textit{DENT} with these synthetic materials, two cracks are forming. These start from both notches, with one arresting after a shorter distance while the other takes a path that is similar to the one observed for toffee.
\begin{figure}[tbp!]
		\centering
        \subfloat[Brittle fracture (\textit{here:} $\urbd = \SI{2500}{mm/min}$, $\vartheta=\SI{25}{\degree C}$)]
        {\includegraphics[scale=0.175,trim=0mm 0mm 0mm 0mm]{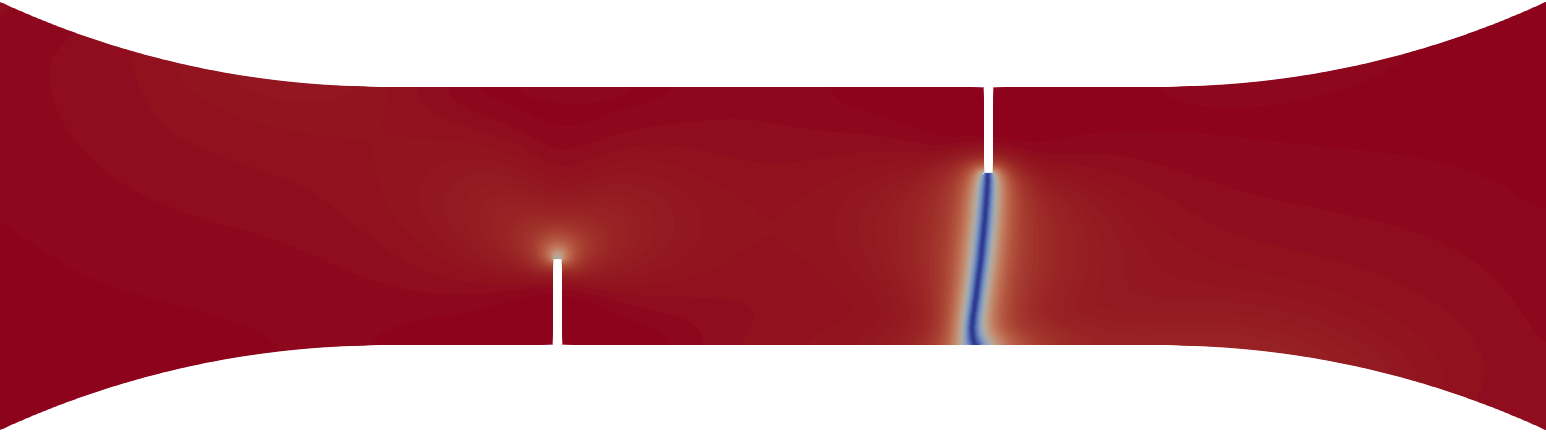}\label{fig:dent_num_brit}} % [trim=left bottom right top]
        
        \vspace{3mm}
		
        \subfloat[Ductile behaviour without crack propagation (\textit{here:} $\urbd = \SI{500}{mm/min}$, $\vartheta=\SI{25}{\degree C}$)]{\includegraphics[scale=0.175,trim=0mm 0mm 0mm 0mm]{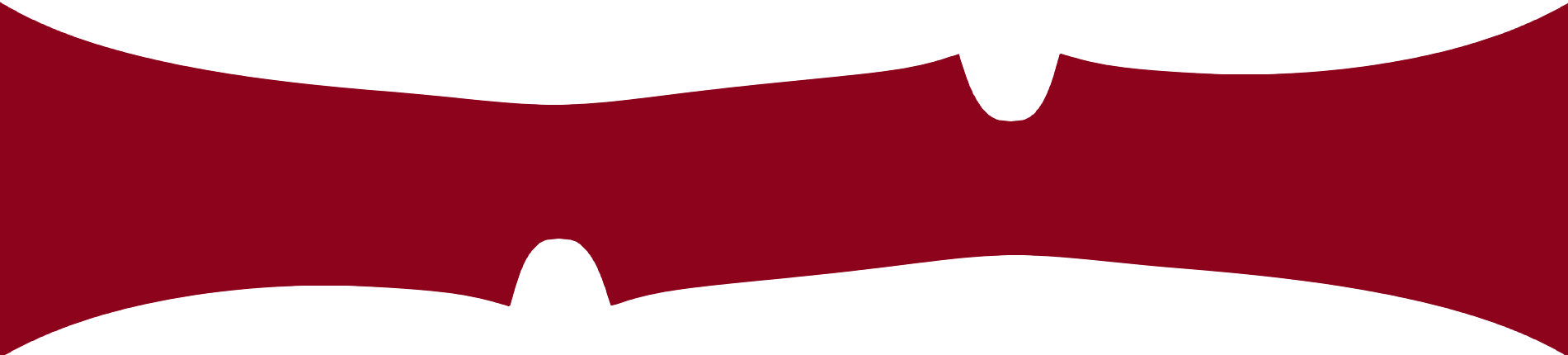}\label{fig:dent_num_duct}} % [trim=left bottom right top]
        
        \vspace{3mm}
        \includegraphics[scale=0.38,trim=0mm 0mm 0mm 0mm]{Figures/Legend_pf.pdf}% [trim=left bottom right top]
		
		\caption{Double-notched specimens under tension with varied displacement rates $\urbd$~(\textit{here:} $\vartheta = \SI{25}{\degree C}$){}\textemdash Numerical results}		
		\label{fig:dent_num}
\end{figure}

The parameterised model reproduces the experimental observations in terms of both $F$-$u$ curves and deformation or crack patterns, Figs.~\ref{fig:dent_exp}--\ref{fig:dent_num}, respectively.
However, when it comes to fracture, if parameters for the rate-dependent fracture resistance $\gc(\te d)$ are prescribed that are constant in space, a second crack of minor length is forming within the specimen. This is similar to the model predictions for elastomers \cite[Figs.~15f.]{dammass2023} and the experimental observations from \cite{han2012}.
Nevertheless, the crack pattern observed experimentally for toffee can be captured by the numerical model as shown in Fig.~\ref{fig:dent_num_brit}. To this end, the fracture resistance parameters, i.e. $\gco$ and $\gct$, are lowered by 5~\% with respect to the remaining domain in the vicinity of one of the notch tips.
Physically, this can be understood as accounting for the slight scatter in the material properties that arise from intrinsic defects and microscopic heterogeneity, which are likely to cause the difference in experimental observations between synthetic and natural materials.

\subsubsection{Model-based analysis}

\begin{figure}[tbp!]
		\centering 
		\subfloat[Brittle fracture\textemdash{}$\dot u=\SI{2500}{mm/min}$]{\includegraphics[width=0.4\linewidth,trim=0mm 0mm 0mm 3mm]{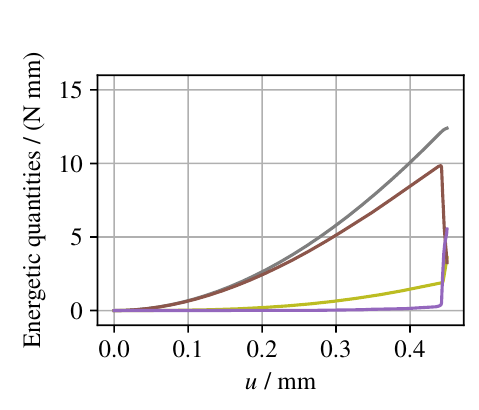}
		\label{fig:sent_ener_brit}} % [trim=left bottom right top]
        \hspace{1.0cm}
        \subfloat[Inelastic deformation without crack propagation\textemdash{}$\dot u=\SI{500}{mm/min}$]{\includegraphics[width=0.425\linewidth,trim=0mm 0mm 0mm -2mm]{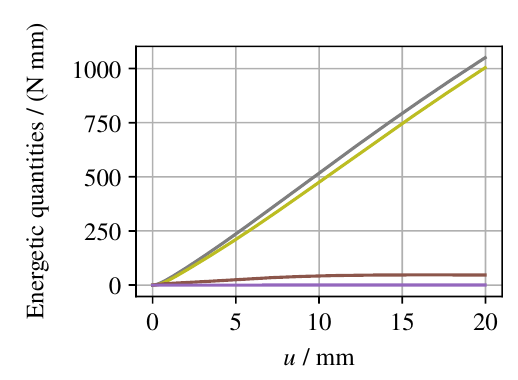}
        \label{fig:sent_ener_duct}} \\% [trim=left bottom right top]
        \includegraphics[width=0.75\linewidth,trim=0mm 5mm 15mm 0mm]{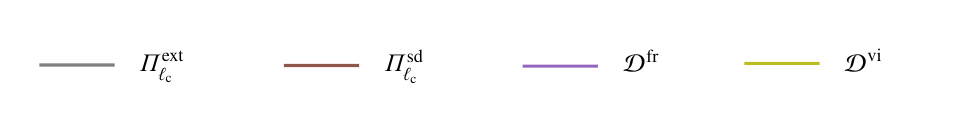}
		\caption{Evolution of integral energetic quantities during \textit{SENT} at room temperature ($\vartheta=\SI{25}{\degree C}$) for brittle fracture at high $\dot u$~(a) and inelastic deformation without crack propagation at low $\dot u$~(b)}
		\label{fig:sent_ener}
\end{figure}
For an analysis of the rate-dependent ductile-to-brittle fracture transition, the evolution of the relevant energetic quantities is discussed in the following.
To this end, the \textit{SENT} is reconsidered. In Fig.~\ref{fig:sent_ener}, for two $\dot u$, the model predictions for the work of external forces supplied to the specimen~$\PiExtlc$ and the integral values of stored free energy~$\Pistlc$ as well as fracture dissipation~$\iDfr$ and viscous dissipation~$\iDvi$ are depicted, which are given by
\begin{equation}
 \PiExtlc = \binte{0}{t}{\Pilcextd}{\tau} \commam
 \Pistlc = \inte{\omref}{\varPsi}{V} \commam
 \iDfr = \binte{0}{t}{\inte{\omref}{\dot D^\nv{fr}}{V}}{\tau} \glmand
 \iDvi = \binte{0}{t}{\inte{\omref}{\dot D^\nv{vi}}{V}}{\tau}
 \comma
\end{equation} 
i.e. the time and volume integrals of the respective rate quantities and densities, respectively.

On the one hand, as it can be seen from Fig.~\ref{fig:sent_ener_brit}, for high $\dot u$, almost all the external work supplied to the specimen~$\PiExtlc$ is stored in terms of strain energy~$\Pistlc$ until the critical load is reached, while the viscous dissipation remains small. 
At that critical point, it suddenly comes to crack propagation, and a significant amount of stored strain energy~$\Pistlc$ is dissipated due to crack propagation, i.e. converted into crack surface pseudo-energy~$\iDfr$. Moreover, the viscous dissipation~$\iDvi$ also raises up, which can be mainly attributed to the zones where no crack propagation takes place.
On the other hand, as shown in Fig.~\ref{fig:sent_ener_duct}, when the \textit{SENT} is performed more slowly, viscous dissipation is dominating the deformation process. Compared to $\iDvi$ and $\PiExtlc$, the stored strain energy $\Pistlc$ remains at a comparably low level and even takes an almost constant value for huge displacements $u \gtrsim \SI{10}{mm}$, which emphasizes the highly dissipative deformation behaviour of the toffee-like caramel under consideration.
Nevertheless, despite $\iDvi \gg \Pistlc$ for low $\dot u$, the amount of strain energy $\Pistlc$ that can be stored is still significantly larger than for high $\dot u$. 
From this, the conclusion can be drawn that the rate-dependent transition between brittle fracture and  significant deformability essentially arises from the complex coupling of a highly dissipative, viscous deformation behaviour and a rate-dependent fracture resistance with the latter competing with the amount of currently stored strain energy, which also depends on the rate of deformation.

\section{Temperature-influence on the mechanical behaviour}
\label{sec:mech-18}

In order to investigate the temperature-dependency, the mechanical behaviour exemplarily is studied for a second constant value of temperature, and the according results are then compared to the aforementioned findings for $\vartheta = \SI{25}{\degree C}$. Here, $\vartheta = \SI{18}{\degree C}$ is chosen.
Similar experiments are conducted as discussed for room temperature in the previous section. Based on these data, another parameterisation of the rate-dependent model is carried out assuming isothermal conditions and following the same procedure as described above.

\subsection{Deformation behaviour}
\begin{figure}[tbp!]
		\centering 
		\includegraphics[width=0.475\linewidth,trim=0mm 0mm 0mm 0mm]{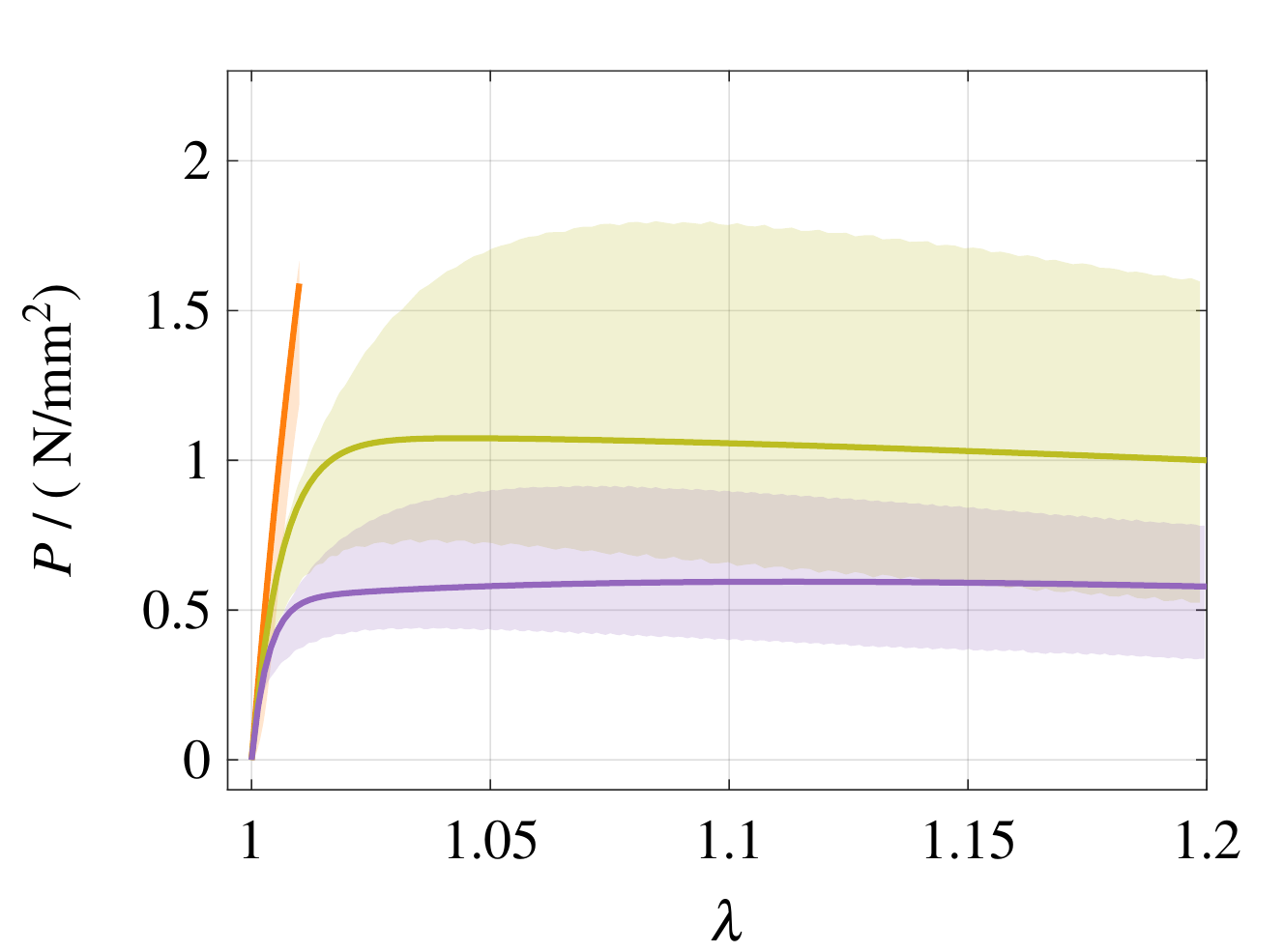} % [trim=left bottom right top]
		\hspace{2mm}
        \includegraphics[scale=0.95,trim=0mm -9mm 0mm 0mm]{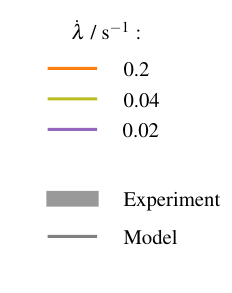}
		\caption{Monotonic tensile tests with unnotched specimens at $\vartheta = \SI{18}{\degree C}$. Experimental stress-stretch data vs. prediction of the parameterised model for different stretch rates}
		
		\label{fig:defo_mon_18C}
\end{figure}

\begin{figure}[tbp!]
		\centering 
		\subfloat[]{\includegraphics[width=0.45\linewidth,trim=0mm 0mm 0mm 5mm]{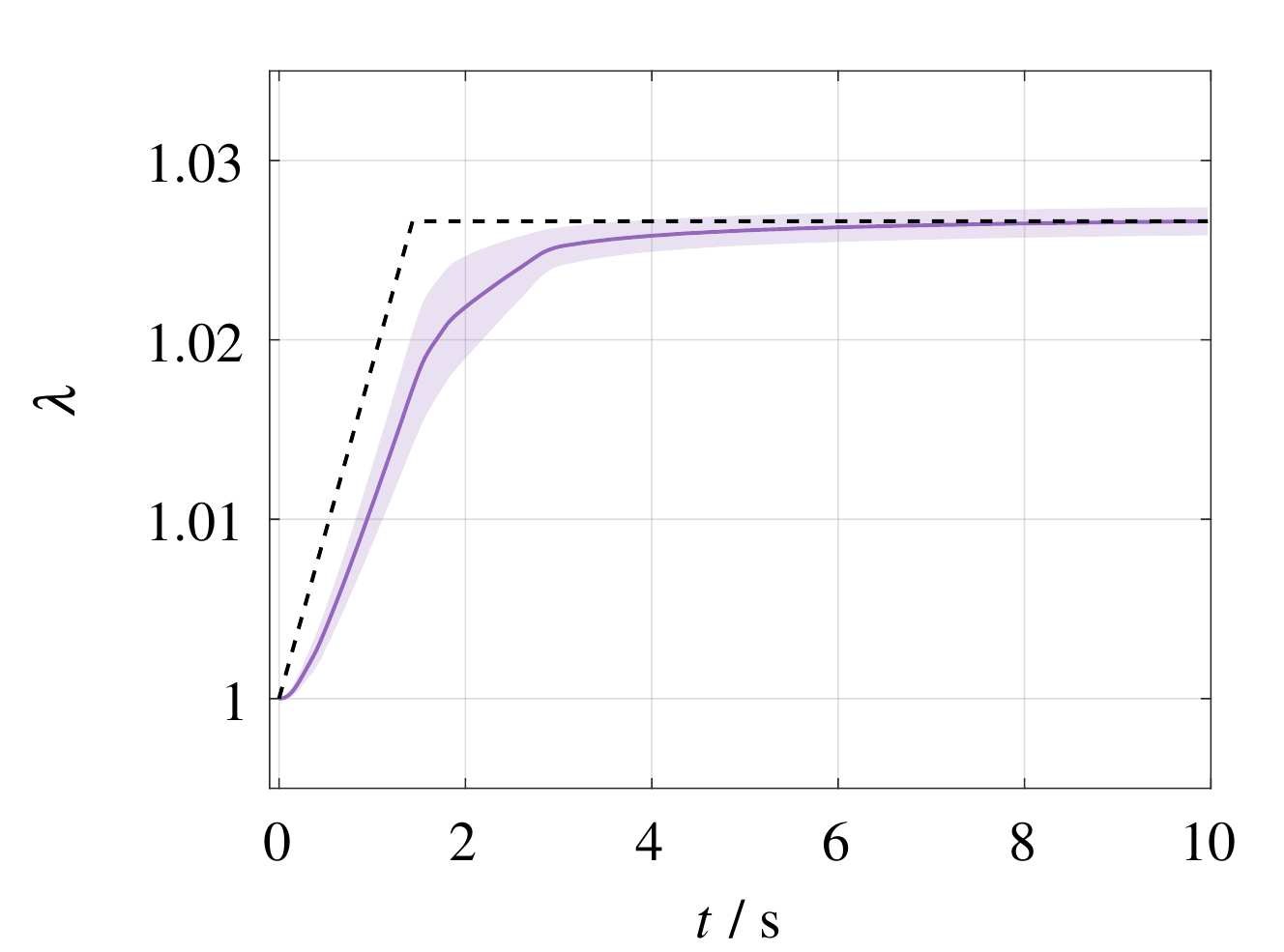}} % [trim=left bottom right top]
		%Prescribed stretch\textemdash{}Actual experimental data vs. idealised curve prescribed to the testing machine
        \hspace{1.0cm}
        \subfloat[]{\includegraphics[width=0.45\linewidth,trim=0mm 0mm 0mm 5mm]{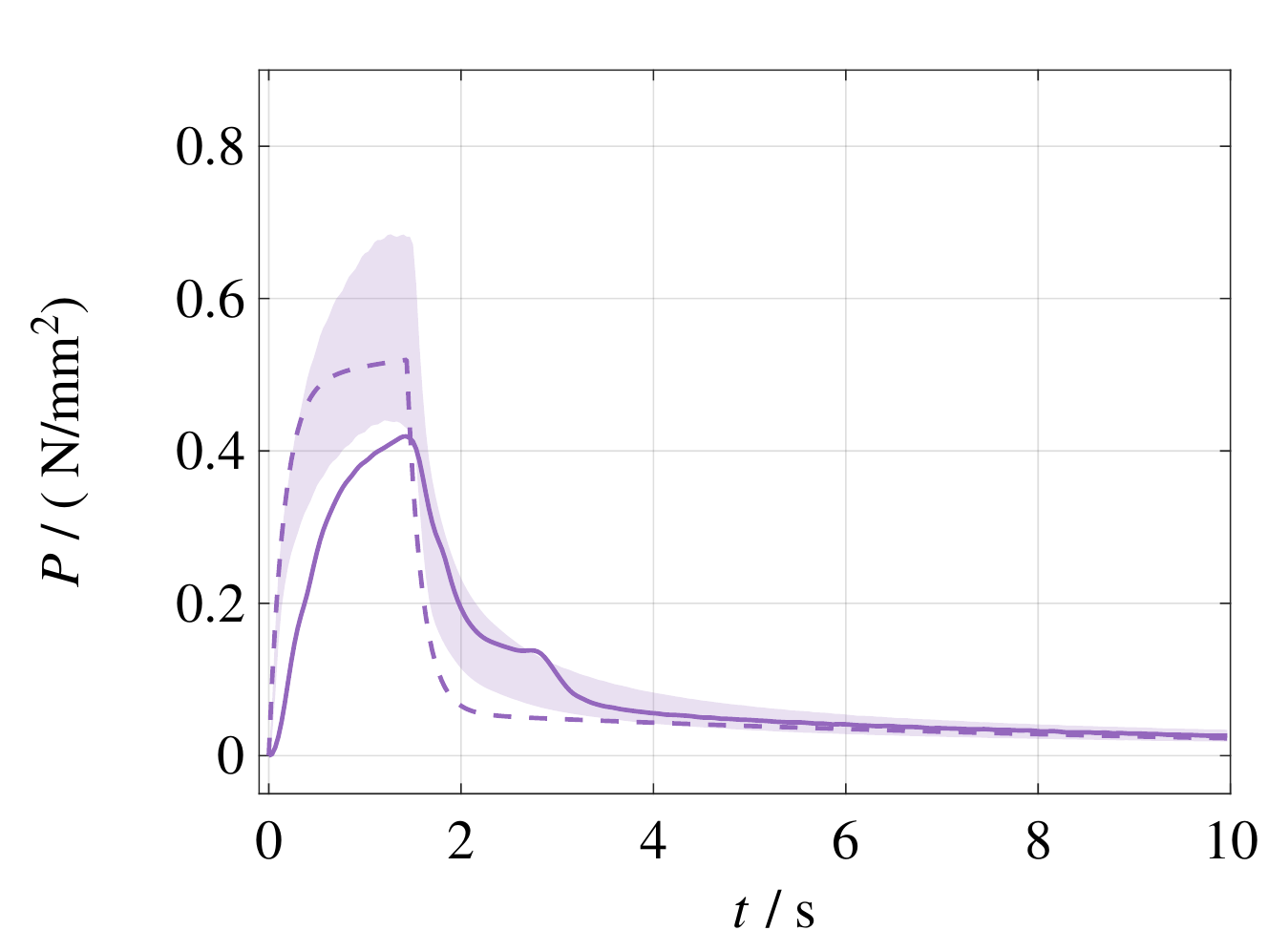} \label{fig:relax_18C_P}} \\% [trim=left bottom right top]
        %[Resulting stress]
        \includegraphics[scale=0.7,trim=0mm 5mm 15mm -2.5mm]{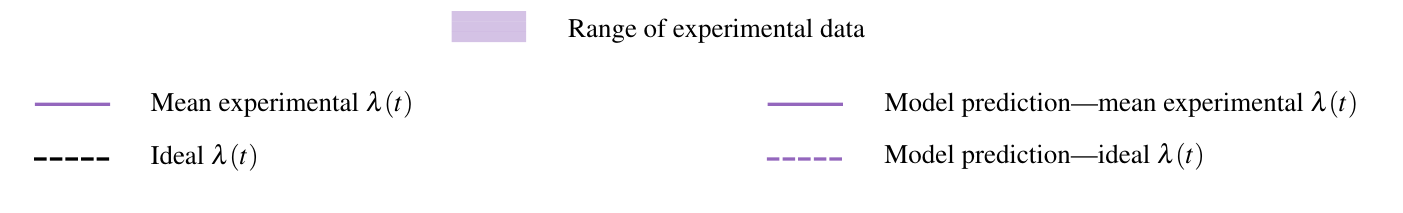}
		\caption{Stress relaxation within unnotched specimens at $\vartheta = \SI{18}{\degree C}$. Actual experimental $\lambda$-$t$ data vs. ideal $\lambda$-$t$ curve (a), and the  experimental $P$-$t$ data compared to the corresponding $P(t)$ predicted by the parameterised model for both the mean experimental $\lambda(t)$ as well as for the ideal $\lambda(t)$ (b)}
		\label{fig:relax_18C}
\end{figure}

For the analysis of the deformation behaviour of the bulk material, unnotched specimens are investigated under monotonic tension at different rates and relaxation conditions as described in Sect~\ref{sec:defo-25}.
The $P$-$\lambda$ curves for monotonic tension at different $\dot \lambda$ are shown in Fig.~\ref{fig:defo_mon_18C}, and relaxation curves are given in Fig.~\ref{fig:relax_18C}. 
Qualitatively, for $\vartheta = \SI{18}{\degree C}$, the rate-dependency of the deformation behaviour is identical to ambiance conditions ($\vartheta = \SI{25}{\degree C}$).
Nevertheless, three consequences of a decrease in temperature become apparent from comparison of Figs.~\ref{fig:defo_mon_18C} and \ref{fig:defo_mon_25C}, and Figs.~\ref{fig:relax_18C_P} and \ref{fig:relax_25C_P}.
Firstly, given a certain value of stretch $\lambda$ in the monotonic experiments, a similar level of stress $P$ is reached for a lower $\dot \lambda$ when temperature is reduced.
Secondly, in the relaxation tests, when temperature is lowered, the material takes considerably more time until the stress does approximately coincide with the equilibrium value, i.e. until $P \approx 0$.
Thirdly, an embrittlement has to be stated. More precisely, when $\dot \lambda$ exceeds a certain value, the specimens fail at very small deformation due to intrinsic defects as it is discussed above, and at $\vartheta = \SI{18}{\degree C}$ this effect can be observed for significantly smaller rates $\dot \lambda$ than when $\vartheta = \SI{25}{\degree C}$.
As a consequence, it can be concluded that there is a qualitative equivalence between a decrease in temperature and an increase in strain rate.


For the number of over-stress branches, Ogden terms and the parameter identification procedure, the statements of Sect.~\ref{sec:defo-25} do likewise apply for $\vartheta = \SI{18}{\degree C}$, and the final parameters are given in Tab.~\ref{tab:matpar-bulk_18}.
From comparison between model and mean experimental data, Figs.~\ref{fig:relax_18C_P} and \ref{fig:defo_mon_18C}, good agreement can be stated, generally.
However, regarding the stress response for the relaxation experiments, there is an artefact in the numerically predicted $P(t)$ when the mean experimental $\lambda(t)$ is considered.
Specifically, there is a sort of a saddle point in the post-peak range of the $P$-$t$~curve, which is not observed in the experiments.
Nevertheless, the numerical $P(t)$ remains in the range of experimental data, and furthermore, such an artefact is not observed when an ideal $\lambda$-$t$~curve is considered as input data to the model instead of the mean experimental $\lambda(t)$, which is additionally shown in Fig.~\ref{fig:relax_18C}.

From comparison of Tabs.~\ref{tab:matpar-bulk_18} and \ref{tab:matpar-bulk_25}, it can also be seen that the relevant relaxation times $\tauX$ are roughly one order of magnitude larger for $\vartheta = \SI{18}{\degree C}$ than for $\vartheta = \SI{25}{\degree C}$. This demonstrates the significance of the influence of temperature on the mechanical behaviour, and that there is an equivalence between the decrease in temperature and an increase in rate of deformation.
Similar observations have also been made employing \textit{Dynamic mechanical analysis} to toffee-like confectionery, see e.g. \cite{schmidt2018}, where a \textit{time-temperature superposition approach} is pursued for toffee.

\begin{table}[tb]
    \caption{Parameters of the model of bulk deformation at $\vartheta = \SI{18}{\degree C}$}
    \centering
    \small
%     \resizebox{\linewidth}{!}{
    \begin{tabular}{*7c}
    \toprule
    $\varXi$ &
    $\nuovnX$ & $\muovX[1]{\varXi} / (\nv{N/mm^2})$ & $\alovX[1]{\varXi}$ & $\muovX[2]{\varXi} / (\nv{N/mm^2})$ & $\alovX[2]{\varXi}$ & $\tauX / \nv{s}$ \\
    \midrule
    1 & \multirow{2}{*}{0.47} & 99771.4 & 0.00129 & 0 & -- & 0.14 \\[3pt]
    2 & & $- 0.285$ & $- 5.93$ & 125.95 & $2.2 \cdot 10^{-6}$& 8.9 \\
        \bottomrule
    \end{tabular}
%     }
    \label{tab:matpar-bulk_18}
\end{table}

\subsection{Fracture behaviour}
\begin{figure}[tbp!]
		\centering 
		{\includegraphics[width=0.45\linewidth,trim=0mm 0mm 0mm 0mm]{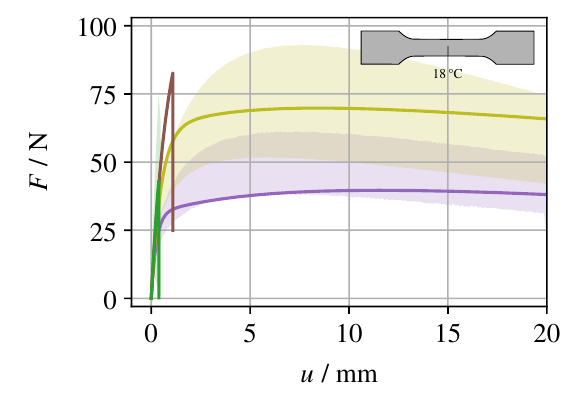}} % [trim=left bottom right top]
		\hspace{2mm}
        \includegraphics[scale=0.9,trim=0mm -15mm 0mm 5mm]{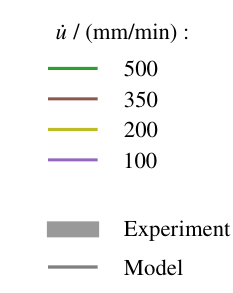}
		\caption{Monotonic tensile tests with single-notched specimens at~$\vartheta = \SI{18}{\degree C}$.
		Experimental force-displacement data compared to the model.
		For $\dot u = \SI{350}{mm/min}$, i.e. for an intermediate displacement rate in the range where the transition between  ductile and highly brittle behaviour, experimentally observed phenomena are ambiguous to a certain extent, which is why the corresponding simulated curve is shown, only.}		
		\label{fig:sent_fu_18C}
\end{figure}

\begin{figure}[tbp!]
		\centering 
		{\includegraphics[width=0.45\linewidth,trim=0mm 0mm 0mm 0mm]{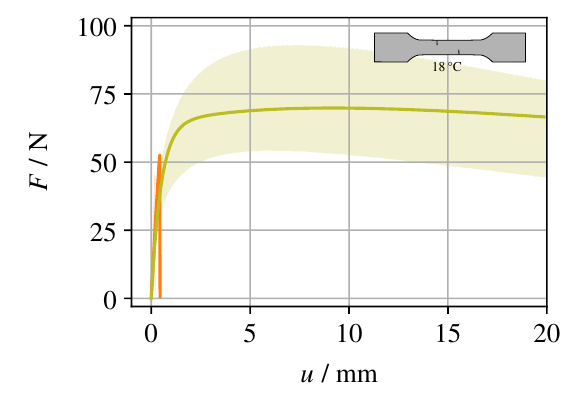}} % [trim=left bottom right top]
		\hspace{2mm}
        \includegraphics[scale=0.9,trim=0mm -12mm 0mm 5mm]{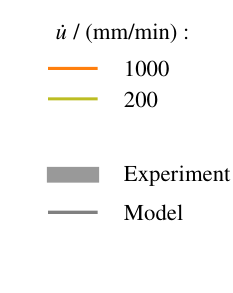}
		\caption{Monotonic tensile tests with double-notched specimens at~$\vartheta = \SI{18}{\degree C}$}		
		\label{fig:dent_fu_18C}
\end{figure}

For the purpose of characterising the temperature influence on the fracture resistance, the experimental procedure described in Sect.~\ref{sec:frac-25} for ambiance conditions is repeated for $\vartheta = \SI{18}{\degree C}$.
Immaterial of temperature, the same three classes of crack or deformation phenomena can be observed, respectively, which are discussed above. Likewise, generally similar $F$-$u$ data are obtained for $\vartheta = \SI{18}{\degree C}$ as for ambiance conditions, see Figs.~\ref{fig:sent_fu_18C} and \ref{fig:dent_fu_18C}.
However, the range of displacement rates $\dot u$ where the transition between ductile behaviour and highly brittle responses appears is shifted to lower values for the lower temperature.

The parameters of the phase-field model, which are identified from the \textit{SENT} similarly to Sect.~\ref{sec:frac-25}, are given in Tab.~\ref{tab:matpar-pf_18}. For these parameters, the model acceptably reproduces the experimental $F$-$u$ data for both the \textit{SENT} and the \textit{DENT}, see Figs.~\ref{fig:sent_fu_18C} and \ref{fig:dent_fu_18C}.
In terms of the rate-dependent fracture resistance $\gc(\te d)$, the decrease in temperature does in particular lead to a decrease of the parameter $\rref$, i.e. scalar equivalent rate of deformation, at which the transition between high and low fracture resistance takes place.
Accordingly, for toffee, there obviously is an equivalence between a decrease in temperature and an increase in strain rate for both the deformation behaviour and the resistance against fracture.
\begin{table}[tb]
    \caption{Parameters of the fracture phase-field for $\vartheta = \SI{18}{\degree C}$}
    \centering
    \small
\begin{tabular}{*7c}
    \toprule
    $\gco / (\nv{N/mm})$ & $\gct / (\nv{N/mm})$ & $c / \nv{s}$ & $\rref / \nv{s}^{-1}$   & $ \etaf / (\nv{N s}/\nv{mm}^2)$  & $k$ & $\lc / \nv{mm}$ \\
    \midrule
    16 & 0.1 & 20 & 1.3 & $5 \cdot 10^ {-6}$ & $10^{-10}$ & 0.25 \\
    \bottomrule
        
\end{tabular}
    \label{tab:matpar-pf_18}
\end{table}

\section{Conclusions and outlook}
\label{sec:concl}

The mechanical behaviour of a toffee-like caramel confectionery is investigated. An experimental study is performed, for which a food material has been produced under lab conditions.
Depending on the rate of deformation, the experimental results differ drastically. Whereas specimens fail in a brittle manner when the strain rate is high, they can be significantly deformed in case of low rates. This phenomenon is referred to as rate-dependent ductile-to-brittle fracture transition.
Furthermore, the temperature-dependency of the mechanical behaviour also is pronounced, and there is an analogy between an increase in rate of deformation and a decrease in temperature.

For the modelling and analysis of these phenomena, a fracture phase-field approach is pursued that combines both, a rate-dependent model of bulk deformation and a rate-dependent Griffith type fracture resistance.
In order to derive the governing equations of the model, an incremental variational principle is presented, which is based on a time-continuous pseudo-potential of rate type.
The parameters are identified in a step-by-step procedure. Whereas the viscous deformation behaviour is characterised exploiting monotonic experiments and relaxation tests with unnotched specimens, the fracture resistance is parameterised in an inverse manner based on experimental data from pre-notched specimens, for what the viscous model of deformation is a priori fixed.
The parameterised model captures the experimental observations.
Moreover, the numerical analysis enables to conclude that it essentially is the complex interplay between the highly dissipative viscous deformation of the bulk material and the rate-dependent fracture resistance that is responsible for the ductile-to-brittle fracture transition.

Exemplarily, the consequences of a decrease in temperature with respect to ambiance conditions are investigated in terms of both experiment and model.
It is shown that there is a qualitative equivalence between a decrease in temperature and an increase in rate of deformation for both the viscous deformation behaviour and the rate-dependent fracture resistance.

The present contribution deals with a food type natural material, and a significant amount of scatter in the experimental data is thus inevitable. In particular, slight deviations in the exact chemical composition and small microscopic heterogeneities or defects are inherent to toffee.
In future work, the present model could be extended to take the stochasticity of the material properties into account. Moreover, the study on the influence of temperature on the mechanical behaviour could be expanded in order to verify if a quantitative \textit{time-temperature equivalence} can be established. For some materials, such relations serve as a basis for \textit{time-temperature superposition principles}, that are typically established considering small deformations and linear viscoelasticity, so far, though.
Finally, the present work may serve as a basis for a physics-based optimisation of industrial food cutting processes. Until now, the approaches that are pursued in the industry are mainly of empirical nature.

% \begin{acknowledgements}
\section*{Acknowledgements}
Support for this research was provided by the German Research Foundation (DFG) under grants \mbox{KA 3309/9-1} and \mbox{RO 3454/7-1}.
The authors gratefully acknowledge Jörg Brummund and Karl A. Kalina for the fruitful discussions.

The computations were performed on a HPC cluster at the Centre for Information Services and High Performance Computing (ZIH) at TU Dresden. The authors thank the ZIH for allocation of computational time. 
% \end{acknowledgements}

\section*{CRediT author contribution statement}
\textit{Franz Dammaß:} Conceptualisation, methodology, investigation -- modelling \& experiments, software, writing.
\textit{Dennis Schab:} Methodology -- material \& specimen preparation, investigation -- experiments.
\textit{Harald Rohm}: Supervision, resources, funding acquisition.
\textit{Markus Kästner:} Conceptualisation, supervision, resources, funding acquisition.

\section*{Declaration of competing interest}
There is no conflict of interest to declare.

% REFERENCES
\bibliographystyle{spphys}       
{\small \bibliography{bibliography.bib}}

\end{document}